\documentclass[longauth]{aa} 

\usepackage{graphicx}
\usepackage{txfonts}
\usepackage{multirow}
\usepackage{colortbl}

\usepackage{hyperref}  
\hypersetup{colorlinks=true,linkcolor=[rgb]{1.,0.2,0.2},citecolor=[rgb]{0.1,0.4,1.},filecolor=[rgb]{0.7,0.2,0.2},urlcolor=[rgb]{0.7,0.2,0.2}}

\usepackage{color}
\definecolor{blue}{rgb}{0., 0., 1}

\newcommand{\oiiiv}{[\textrm{O}\textsc{iii}]\ensuremath{\lambda5007}}
\newcommand{\oiiiiv}{[\textrm{O}\textsc{iii}]\ensuremath{\lambda4959}}

\newcommand{\ha}{\ifmmode {\rm H}\alpha \else H$\alpha$\fi}
\newcommand{\halam}{\ifmmode {\rm H}\alpha \lambda6563 \else H$\alpha$ $\lambda$6563 \fi}
\newcommand{\hb}{\ifmmode {\rm H}\beta \else H$\beta$\fi}
\newcommand{\hg}{\ifmmode {\rm H}\gamma \else H$\gamma$\fi}
\newcommand{\hblam}{\ifmmode {\rm H}\beta \lambda4861 \else H$\beta$ $\lambda$4861 \fi}
\newcommand{\lya}{\ifmmode {\rm Ly}\alpha \else Ly$\alpha$\fi}
\newcommand{\pg}{\ifmmode {\rm P}\gamma \else Pa$\gamma$\fi}
\newcommand{\lyb}{\ifmmode {\rm Ly}\beta \else Ly$\beta$\fi}
\newcommand{\lyg}{\ifmmode {\rm Ly}\gamma \else Ly$\gamma$\fi}

\newcommand{\flyc}{\ifmmode  \mathrm{f}_\mathrm{esc}\mathrm{(LyC)} \else $\mathrm{f}_\mathrm{esc}\mathrm{(LyC)}$\fi}

\def\ergs{\ifmmode \mathrm{erg\hspace{1mm}s}^{-1} \else erg s$^{-1}$\fi}

\def\micron{\ifmmode \mu\mathrm{m} \else $\mu$m\fi}
\def\msun{\ifmmode \mathrm{M}_{\odot} \else M$_{\odot}$\fi}
\def\msunyr{\ifmmode \mathrm{M}_{\odot} \hspace{1mm}{\rm yr}^{-1} \else $\mathrm{M}_{\odot}$ yr$^{-1}$\fi}
\def\zsun{\ifmmode Z_{\odot} \else Z$_{\odot}$\fi}
\def\lsun{\ifmmode L_{\odot} \else L$_{\odot}$\fi}
\def\mstar{\ifmmode \mathrm{M}_{\star} \else M$_{\star}$\fi}
\newcommand{\JWST}{\textrm{JWST}}
\newcommand{\HST}{\textrm{HST}}

\usepackage{orcidlink}

\newcommand{\orcid}[1]{\href{https://orcid.org/#1}{\textcolor[HTML]{A6CE39}{\aiOrcid}}}

\begin{document}

\titlerunning{\JWST\ probes star clusters at $z\simeq10$}

\title{The $z = 9.625$ Cosmic Gems Galaxy was a ``Compact Blue Monster'' Propelled by Massive Star Clusters \thanks{Based on observations collected with the \textit{James Webb} Space Telescope (\JWST) and \textit{Hubble} Space Telescope (\HST).
These observations are associated with \JWST\ program n.4212 (PI L. Bradley) and program n. 5917 (PI E. Vanzella).}}

\authorrunning{Eros Vanzella et al.}
\author{
E.~Vanzella\inst{\ref{inafbo}}\fnmsep\thanks{E-mail: \href{mailto:eros.vanzella@inaf.it}{eros.vanzella@inaf.it}}$^{\orcidlink{0000-0002-5057-135X}}$ \and
M.~Messa\inst{\ref{inafbo}}$^{\orcidlink{0000-0003-1427-2456}}$ \and
A.~Adamo\inst{\ref{univstock}}$^{\orcidlink{0000-0002-8192-8091}}$\and
F.~Loiacono\inst{\ref{inafbo}}$^{\orcidlink{0000-0002-8858-6784}}$ \and
M. Oguri\inst{\ref{ChibaUniversity},\ref{chibaphys}}$^{\orcidlink{0000-0003-3484-399X}}$ \and
K.~Sharon\inst{\ref{um}}$^{\orcidlink{0000-0002-7559-0864}}$  \and 
L.~D.~Bradley\inst{\ref{STScI}}$^{\orcidlink{0000-0002-7908-9284}}$ \and
P.~Bergamini\inst{\ref{unimi},\ref{inafbo}}$^{\orcidlink{0000-0003-1383-9414}}$ \and
M.~Meneghetti\inst{\ref{inafbo}}$^{\orcidlink{0000-0003-1225-7084}}$\and 
A.~Claeyssens\inst{\ref{CRAL}}$^{\orcidlink{0000-0001-7940-1816}}$  \and
B.~Welch\inst{\ref{Goddard},\ref{univmaryland}}$^{\orcidlink{0000-0003-1815-0114}}$\and
M.~Brada\v{c} \inst{\ref{uniLjubljana}}$^{\orcidlink{0000-0001-5984-0395}}$ \and 
A.~Zanella\inst{\ref{inafbo}}$^{\orcidlink{0000-0001-8600-7008}}$ \and
A.~Bolamperti\inst{\ref{mpa},\ref{inafiasf}}$^{\orcidlink{0000-0001-5976-9728}}$\and
F.~Calura\inst{\ref{inafbo}}$^{\orcidlink{0000-0002-6175-0871}}$\and
T.~Y-Y.~Hsiao\inst{\ref{CfA},\ref{JHU},\ref{STScI}}$^{\orcidlink{0000-0003-4512-8705}}$ \and
E.~Zackrisson\inst{\ref{UU}}$^{\orcidlink{0000-0003-1096-2636}}$ \and
M.~Ricotti\inst{\ref{univmaryland}}$^{\orcidlink{0000-0003-4223-7324}}$ \and
L.~Christensen\inst{\ref{univcope},\ref{dawn}}$^{\orcidlink{0000-0001-8415-7547}}$ \and
J.~M.~Diego\inst{\ref{ifca}}$^{\orcidlink{0000-0001-9065-3926}}$ \and
F.~E.~Bauer\inst{\ref{Tarapaca}}$^{\orcidlink{0000-0002-8686-8737}}$ \and 
X.~Xu\inst{\ref{CIREA}}$^{\orcidlink{0000-0000-0000-0001}}$  \and
S.~Fujimoto\inst{\ref{uoft},\ref{dunlap}}$^{\orcidlink{0000-0001-7201-5066}}$ \and 
C.~Grillo \inst{\ref{unimi},\ref{inafiasf}}$^{\orcidlink{0000-0002-5926-7143}}$\and
M.~Lombardi\inst{\ref{unimi}}$^{\orcidlink{0000-0002-3336-4965}}$ \and
P.~Rosati \inst{\ref{unife},\ref{inafbo}}$^{\orcidlink{0000-0002-6813-0632}}$\and
T.~Resseguier\inst{\ref{STScI},\ref{JHU}}$^{\orcidlink{0009-0007-0522-7326}}$\and
A.~Zitrin\inst{\ref{BGU}} $^{\orcidlink{0000-0002-0350-4488}}$ \and
A.~Bik\inst{\ref{univstock}}$^{\orcidlink{0000-0001-8068-0891}}$\and
J.~Richard\inst{\ref{CRAL}}$^{\orcidlink{0000-0001-5492-1049}}$\and
Abdurro'uf\inst{\ref{JHU},\ref{STScI}}$^{\orcidlink{0000-0002-5258-8761}}$ \and
R.~Bhatawdekar\inst{\ref{ESAC_spain}}$^{\orcidlink{0000-0003-0883-2226}}$ \and
D.~Coe\inst{\ref{STScI}}$^{\orcidlink{0000-0001-7410-7669}}$ \and
B.~Frye\inst{\ref{UniversityArizona}}$^{\orcidlink{0000-0003-1625-8009}}$ \and 
A.~K.~Inoue\inst{\ref{Waseda1},\ref{Waseda2}}$^{\orcidlink{0000-0002-7779-8677}}$ \and
Y.~Jimenez-Teja\inst{\ref{granada},\ref{ON}}$^{\orcidlink{0000-0002-6090-2853}}$ \and
C.~Norman\inst{\ref{JHU}, \ref{STScI}}$^{\orcidlink{0000-0002-5222-5717}}$ \and
J.~R.~Rigby\inst{\ref{Goddard}}$^{\orcidlink{0000-0002-7627-6551}}$ \and  
M.~Trenti\inst{\ref{unimelb}}$^{\orcidlink{0000-0001-9391-305X}}$ \and
T.~Hashimoto\inst{\ref{UniversityofTsukuba}}$^{\orcidlink{0000-0002-0898-4038}}$
}
\institute{
INAF -- OAS, Osservatorio di Astrofisica e Scienza dello Spazio di Bologna, via Gobetti 93/3, I-40129 Bologna, Italy \label{inafbo} 
\and
Department of Astronomy, Oskar Klein Centre, Stockholm University, AlbaNova University Center, SE-106 91, Sweden\label{univstock}
\and
Center for Frontier Science, Chiba University, 1-33 Yayoi-cho, Inage-ku, Chiba 263-8522, Japan\label{ChibaUniversity}
\and
Department of Physics, Graduate School of Science, Chiba University, 1-33 Yayoi-Cho, Inage-Ku, Chiba 263-8522, Japan \label{chibaphys}
\and
Department of Astronomy, University of Michigan 1085 South University Avenue Ann Arbor, MI 48109, USA\label{um}
\and
Space Telescope Science Institute (STScI), 3700 San Martin Drive, Baltimore, MD 21218, USA \label{STScI}
\and
Dipartimento di Fisica, Università degli Studi di Milano, Via Celoria 16, I-20133 Milano, Italy\label{unimi}
\and
Univ Lyon, Univ Lyon1, ENS de Lyon, CNRS, Centre de Recherche Astrophysique de Lyon UMR5574, Saint-Genis-Laval, France\label{CRAL}
\and
Astrophysics Science Division, Code 660, NASA Goddard Space Flight Center, 8800 Greenbelt Rd., Greenbelt, MD, 20771, USA \label{Goddard}
\and
Department of Astronomy, University of Maryland, College Park, 20742, USA\label{univmaryland}
\and
University of Ljubljana, Faculty of Mathematics and Physics, Jadranska ulica 19, SI-1000 Ljubljana, Slovenia\label{uniLjubljana}
\and
Max-Planck-Institut f\"ur Astrophysik, Karl-Schwarzschild-Str. 1, D-85748 Garching, Germany \label{mpa}
\and 
INAF -- IASF Milano, via A. Corti 12, I-20133 Milano, Italy\label{inafiasf}
\and
Center for Astrophysics \text{\textbar} Harvard \& Smithsonian, 60 Garden Street, Cambridge, MA 02138, USA \label{CfA}
\and
Center for Astrophysical Sciences, Department of Physics and Astronomy, The Johns Hopkins University, 3400 N Charles St. Baltimore, MD 21218, USA \label{JHU}
\and
Observational Astrophysics, Department of Physics and Astronomy, Uppsala University, Box 516, SE-751 20 Uppsala, Sweden \label{UU}
\and
Niels Bohr Institute, University of Copenhagen, Jagtvej 128, DK-2200 Copenhagen N, Denmark\label{univcope}
\and
Cosmic Dawn Center (DAWN), Denmark\label{dawn}
\and
 Instituto de F\'isica de Cantabria (CSIC-UC). Avda. Los Castros s/n. 39005 Santander, Spain \label{ifca}
\and
Instituto de Alta Investigaci{\'{o}}n, Universidad de Tarapac{\'{a}}, Casilla 7D, Arica, Chile \label{Tarapaca}
\and
Center for Interdisciplinary Exploration and Research in Astrophysics (CIERA), 1800 Sherman Avenue, Evanston, IL, 60201, USA\label{CIREA}
\and
David A. Dunlap Department of Astronomy and Astrophysics, University of Toronto, 50 St. George Street, Toronto, Ontario, M5S 3H4, Canada \label{uoft}
\and
Dunlap Institute for Astronomy and Astrophysics, 50 St. George Street, Toronto, Ontario, M5S 3H4, Canada \label{dunlap} 
\and
Dipartimento di Fisica e Scienze della Terra, Università degli Studi di Ferrara, Via Saragat 1, I-44122 Ferrara, Italy\label{unife}
\and 
Department of Physics, Ben-Gurion University of the Negev, P.O. Box 653, Be'er-Sheva 84105, Israel\label{BGU}
\and
European Space Agency (ESA), European Space Astronomy Centre (ESAC), Camino Bajo del Castillo s/n, 28692 Villanueva de la Cañada, Madrid, Spain \label{ESAC_spain}
\and
Department of Astronomy/Steward Observatory, University of Arizona, 933 N. Cherry Avenue, Tucson, AZ 85721, USA \label{UniversityArizona}
\and
Department of Physics, School of Advanced Science and Engineering, Faculty of Science and Engineering, Waseda University, 3-4-1, Okubo, Shinjuku, Tokyo 169-8555, Japan\label{Waseda1}
\and
Waseda Research Institute for Science and Engineering, Faculty of Science and Engineering, Waseda University, 3-4-1, Okubo, Shinjuku, Tokyo, 169-8555, Japan\label{Waseda2}
\and
Instituto de Astrofísica de Andalucía–CSIC, Glorieta de la Astronomía s/n, E-18008, Granada, Spain \label{granada}
\and
Observatório Nacional – MCTI (ON), Rua General José Cristino, 77, São Cristóvão, 20921-400, Rio de Janeiro, Brazil \label{ON}
\and
School of Physics, The University of Melbourne, VIC, 3010, Australia\label{unimelb}
\and
Division of Physics, Faculty of Pure and Applied Sciences, University of Tsukuba, Tsukuba, Ibaraki 305-8571, Japan \label{UniversityofTsukuba}
}

\date{} 

\abstract
{
The recent discovery of five massive stellar clusters at $z= 9.625$ in the {\it Cosmic Gems} has raised the question about the formation mechanism of star clusters in the first half Gyr after the Big-Bang. We infer the total stellar mass in clusters by normalizing and integrating the stellar cluster mass function (SCMF, $dn(M)/dM~=~n_{0}M^{\beta}$), assuming three different slopes $\beta = -1.5$, $-$2.0 and $-$2.5 and different lower-mass limits between $10^2$ and $10^5$~\msun. 
The total integrated cluster stellar mass is compared to the stellar mass inferred from the counter-image of the Cosmic Gems, which provides the best, modestly magnified ($\mu = 1.84\pm0.05$) representation of the entire galaxy. The delensed stellar mass of the Cosmic Gems galaxy is estimated as $3.5_{-1.8}^{+3.3}\times 10^7$ \msun, with an effective radius of R$_{\rm eff} = 
103_{-15}^{+13}$ parsec and a stellar surface mass density of $\Sigma_{\rm mass} =  520_{-225}^{+340}$~\msun~pc$^{-2}$.
Accounting for normalization uncertainties — including different lensing magnification scenarios for the arc — a modified SCMF, combined with a significantly high star cluster formation efficiency (approaching 100\%), appears to be a necessary condition to explain the relatively short formation timescale of both the star clusters and the counter-image, without exceeding the galaxy's stellar mass.
By extrapolating the physical properties at the peak of the burst we find that in its recent past ($\lesssim 30$ Myr) the Cosmic Gems galaxy has likely experienced a specific star formation rate (sSFR) exceeding 25~Gyr$^{-1}$ and luminosity approaching the ``blue monster'' regime (M$_{\rm UV} < -20$). Our study provides insights into the extreme clustered nature of star formation in early galaxies and shed light into the formation of  bound star clusters that might survive to $z = 0$ as globular clusters, older than 13 Gyr.
}

   \maketitle

\section{Introduction}
\label{sect:intro}

In the local universe, star formation is hierarchically in a clustered fashion, with bound star clusters being only a fraction of the total stellar mass formed in a given time, typically referred to as cluster formation efficiency, $\Gamma$ \citep[e.g.,][]{krumholz2019}. Bound star cluster populations follow a power-law distribution of their masses, and possibly show a cut-off at the high-mass end, describing the maximum star cluster mass a galaxy might form \citep{adamo2020SSRv}.
The quantity $\Gamma$ has been extensively measured in the local Universe and found to positively correlate with the star formation rate surface density ($\Sigma_{\rm SFR}$) of the main galaxy \citep[e.g.,][]{adamo15, Johnson2016gamma, messa2018}, although it remains difficult to disentangle biases in the methodology, leading  to contrasting results \citep{cook2023}. The observed increase of a galaxy's $\Sigma_{\rm SFR}$ with redshift \citep[e.g.,][]{ormerod2024_sizes, morishita2024_sizes} 
suggests that $\Gamma$ was high in the early universe. However, a direct measure of $\Gamma$ at cosmological distances is extremely challenging and will require extreme adaptive optics with PSF sizes of $\simeq 10-20$ milliarcsec from the extremely large telescope or next-generation instruments on 8m class telescopes coupled with gravitational lensing \citep[see the discussion in][]{vanz22_CFE}. First attempts of measuring $\Gamma$ at high redshift come from exceptionally magnified systems \citep[e.g.,][]{vanz_sunburst}, in which several star clusters have been identified and related to the bursty star formation events in the host galaxy (or even quantifying their contribution to the total ultraviolet light of the host, \citealt{vanzella2_sunrise2023, adamo2024, Messa_D1T1_2025, Mowla2024, Bradac2024, Fujimoto_2024_grapes}).  
Another aspect suggesting a high occurrence of bound star clusters in the early Universe is that high-density conditions on average favor high $\Gamma$, along with the presence of very massive star clusters \citep[e.g.,][]{Garcia2023, Sugimura2024, Kruijssen2025}.

High redshift star clusters were already identified in the pre-JWST era with Hubble \citep[e.g.,][]{vanz_paving, vanz_id14, vanz19, vanz22_CFE} along with several parsec-scale star complexes \citep[e.g.,][]{bouwens_tiny17, rigby17, johnson17, Mestric22, Welch22_clumps}. 
The advent of \JWST\ enabled the identification of similar parsec-scale stellar clump regions with  lower magnifications and/or higher redshift \citep[e.g.,][]{Messa_D1T1_2025, claeyssens24, vanzella2_sunrise2023, Mowla2024, Hsiao2023, Fujimoto_2024_grapes}, and even allowing us to detect relatively old star clusters by means of the extended (NIRCam and MIRI) wavelength range \citep[e.g.,][]{adamo_sparkelr_2023}. 

\begin{figure}
\center
 \includegraphics[width=\columnwidth]{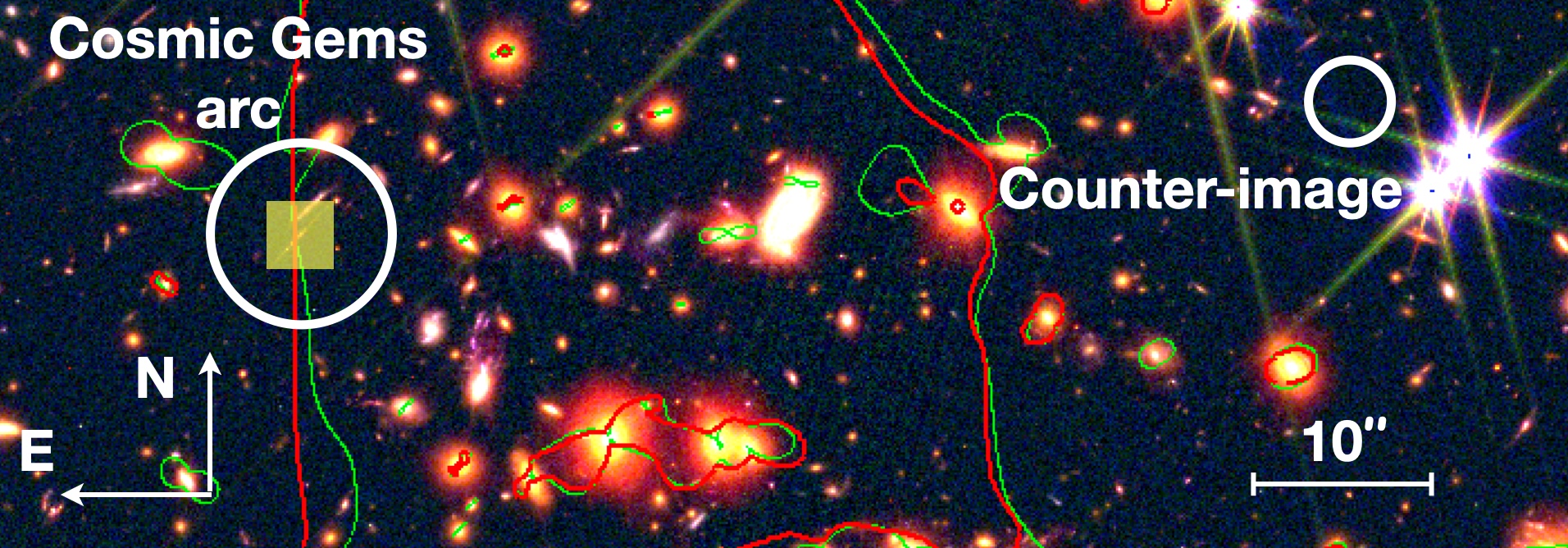}
 \caption{\JWST/ NIRCam color image of the portion of the galaxy cluster SPT0615 field including the CG arc and the CI, (white circles), along with the critical lines at the $z=9.625$ for {\sc Lenstool} \citep{jullo2007} and {\sc glafic} \citep{oguri2010,oguri2021} lens models, marked with red and green lines, respectively. The yellow shaded square marks the field of view of the \JWST/NIRSpec IFU observations (see \citealt[][]{Messa25_CG}).} 
 \label{panoramic}
\end{figure}
Remarkably, the discovery of the super magnified {\it Cosmic Gems} (CG hereafter) arc \citep[][hereafter LB25]{Bradley2025}, now spectroscopically confirmed to be at redshift  $9.625$ \citep[][MM25 in the following]{Messa25_CG} is the first concrete opportunity we have of addressing the internal properties of a galaxy within the first 500 Myr of cosmic time. 
Furthermore, the identification of five very dense ($\sim 10^{5-6}$ \msun~pc$^{-2}$) and massive ($\sim [1-3] \times 10^{6}$ \msun) gravitationally bound star clusters all located within a relatively small physical region of $\sim 70$~pc in the CG galaxy was unexpected \citep[][hereafter AA24]{adamo2024}, and, to our knowledge, has no analogous examples in the local Universe.

In this work we infer the fraction of stellar mass of the CG galaxy which formed in bound stellar clusters. 
We explore the amount of stellar mass located in star clusters by properly integrating the star cluster mass function (SCMF, after varying the slope and the minimum cluster mass) and compare it with the physical properties and stellar mass of the entire galaxy, inferred from \JWST/NIRCam photometric data of the candidate counter-image (dubbed CI hereafter) which offers the best representation of the global properties of the CG galaxy. 

Throughout this paper, we assume a flat cosmology with $\Omega_{M}$= 0.31,
$\Omega_{\Lambda}$= 0.69, and $H_{0} = 67.7\,{\rm km}\,{\rm s}^{-1}\,{\rm Mpc}^{-1}$ \citep[][]{Plank18}, corresponding to 4360 parsec per arcsecond at $z=9.625$. 
All magnitudes are given in the AB system \citep{Oke_1983}:
$m_{\rm AB} = 23.9 - 2.5 \log(f_\nu / \mu{\rm Jy})$.

\begin{figure*}
\center
 \includegraphics[width=\textwidth]{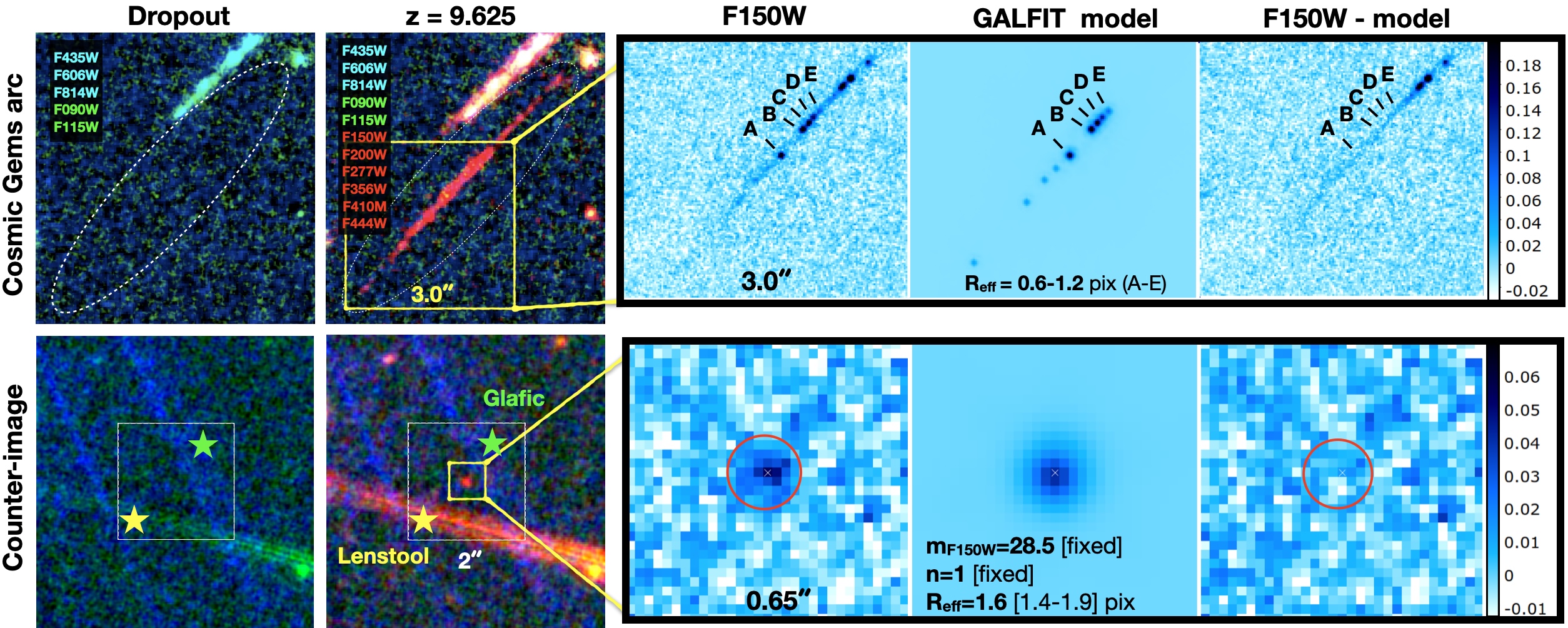}
 \caption{\JWST/ NIRCam imaging and {\tt Galfit} fitting of the Cosmic Gems arc and its counter-image. In the left panels, the sharp dropout of the arc and counter-image in the Hubble (F435W+F606+F814W, blue channel) and \JWST/NIRCam (F090W+F115W, green channel) RGB rendering, with the detection in the redder \JWST/NIRCam bands (red channel of the RGB rendering, as indicated in the figure). In the color image showing the counter-image, the predicted position from the new {\tt Glafic} (yellow star) and {\tt LENSTOOL} (green star) models are reported (from MM25). The regions outlined with yellow squares are zoomed in on the right panels in the NIRCam F150W band, along with the {\tt Galfit} modeling and residuals (in counts units, rightmost panels).} 
 \label{galfit}
\end{figure*}
%

\section{The Cosmic Gems galaxy} \label{GEMS}

\subsection{Star clusters and lensing magnification} 
\label{general}
Initially discovered in the Hubble data from the RELICS survey \citep[][see also \citealt{Welch22_clumps}]{Coe_2019}, the CG arc 
shows (at least) five gravitationally-bound parsec-scale star clusters in \JWST/NIRCam imaging, now confirmed (after 
spectroscopic redshift determination) to be remarkably dense ($\sim 10^{5-6}$ \msun~pc$^{-2}$), massive ($\sim [1-6] \times 
10^{6}$ \msun), with ages spanning the range $8 - 27$ Myr (MM25), and likely located within a magnified portion of the CG galaxy 
smaller than 70 parsec (AA24). 
The initial lens models of the galaxy cluster SPT0615 \citep{adamo2024, Bradley2025} were recently refined by including additional new multiple systems at $2<z<6$ from VLT/MUSE (PI. Bauer) confirming the critical line crossing the arc and very large magnification factors along the arc ($\mu > \times 50-300$; MM25). The improved models also confirm the previous predictions of the position of a CI, which was later detected in the NIRCam images (with S/N~$\simeq 15$), showing the same colors and a very solid photometric redshift compatible to the main arc (LB25, and see Figure~\ref{galfit}). The angular separation between the improved lens model predictions and the candidate CI are $\lesssim 1''$. No other $z\sim 10$ candidates are detected (at the available depth) within a region of $10''$ size centered on the predicted positions. 
Hereafter, we consider this CI as the best proxy of the entire CG galaxy. Such an image is far from any $z=9.625$ critical lines and lies within a few (or fraction of) arcsec from the predicted positions of the current best models (MM25).
The low magnification and small error, $\mu_{host} = 1.84 \pm 0.05$, associated to the CI allow us to derive the global physical properties of the CG galaxy with low lensing-related uncertainty dependence. The very low magnification, however, prevents us from identifying the massive star clusters that we detect in the arc.

In the following analysis, we adopt the updated lensing model predictions for both the CI, $\mu_{host}$ as reported above, and the star clusters, using the updated $\mu_{arc}$ from MM25 (reported in Appendix~\ref{mass_mu}), which closely resemble those published by AA24. Further considerations on how the change in the predicted magnifications affects our results are presented in Appendix~\ref{mass_mu}.

\subsection{Physical properties and morphology of the host galaxy}
\label{physical_properties}

MM25 derived the physical properties of the CG arc by performing JWST/NIRSpec-IFU-based spectral and SED fitting across sub-regions of the arc. The sum of the intrinsic stellar masses inferred for each region yields an estimate of the galaxy's stellar mass, $ \approx 3.5\times 10^7$ \msun. This value is consistent with that derived from the CI (see below), as are the inferred stellar age ($10-40$ Myr) and low dust attenuation (A$_{\rm V}\lesssim 0.2$). Combined with the weak rest-frame optical lines detected in the NIRSpec/IFU datacube $-$ namely, \hb\ and  \oiiiiv\ \footnote{The other component of the doublet, \oiiiv\ (5008.24\AA\ vacuum) is out from the observed spectral range.} (4960.30\AA, vacuum) with equivalent widths of $\simeq20$~\AA\ and $\simeq50$~\AA, respectively $-$ the galaxy appears to be in a currently ‘dormant’ phase of star formation.
As noted by MM25 and analyzed in detail by Christensen et al. in prep., the NIRSpec spectrum exhibits a pronounced \lya\ damping wing that depresses the F150W flux, especially in the arc and its CI (showing relatively high signal-to-noise ratios on NIRCam data). This flux deficit likely caused the previously overestimated photometric redshifts ($z_{\rm phot} > 10$) for both images, compared with the new spectroscopic value $z_{\rm spec}=9.625$.
MM25 also reported the revised physical properties of the star clusters using the photometry extracted by AA24, assuming star formation histories of $\tau=1$ Myr using BPASS models and updating the redshift to the new value. 
Stellar masses in the range $(1-6)\times 10^{6}$ \msun\ and ages $8-27$ Myr were found and agree within the uncertainties with the previous estimates by AA24.

\begin{figure*}
\center
\includegraphics[width=\textwidth]{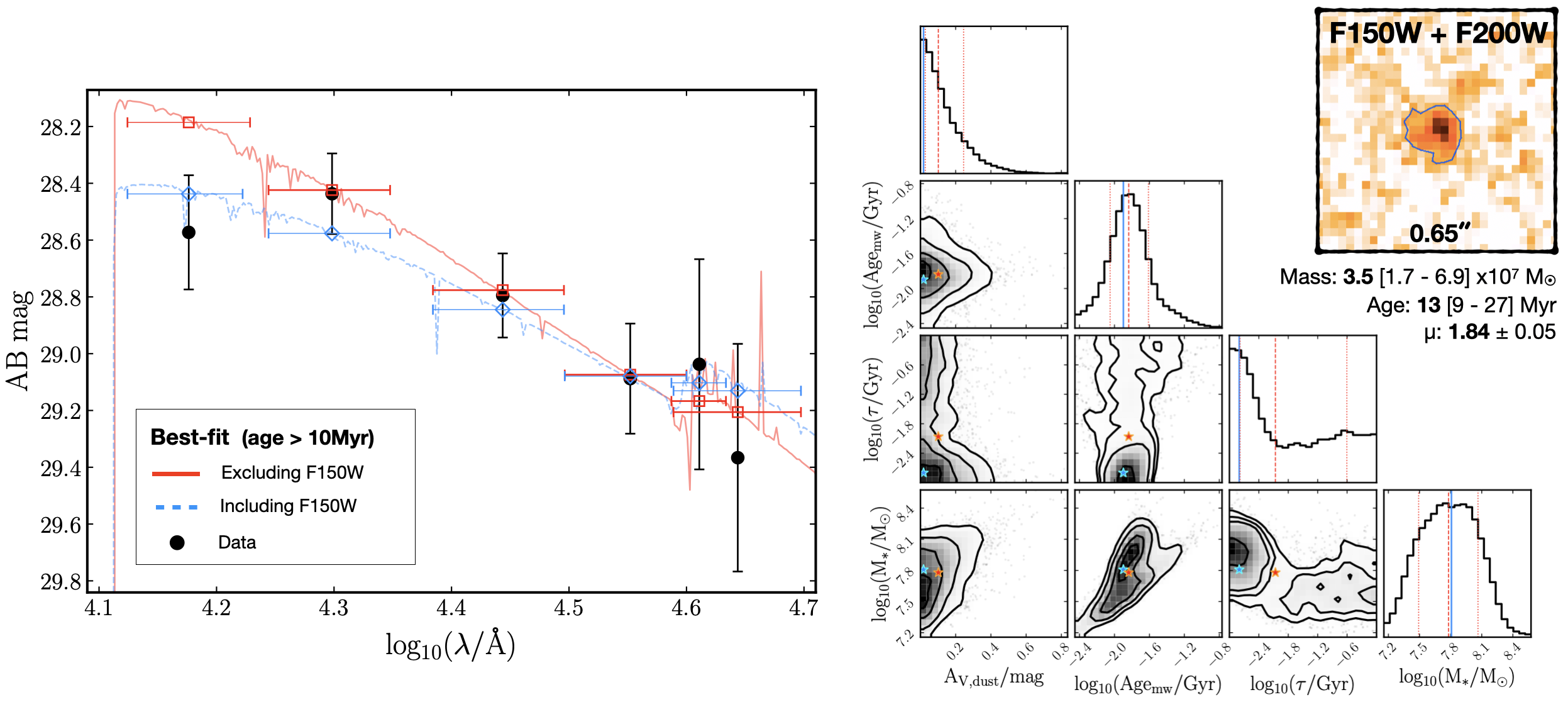}
 \caption{The corner plot (right) and the SED fitting results of the counter-image (left). 
 We fix $z=9.625$ throughout and adopt an age not younger than 10 Myr (see text for more details). Red/blue lines on the left indicate fits excluding/including the F150W data point.
 The red line shows the fiducial SED fit solution (see Figure~\ref{noF150W} for a comparison of corner plots with and without the F150W band data point). Horizontal bars indicate the bandwidth. On the right panel the inset shows the stacked short-wavelength bands (F150W + F200W) image of the CI where the 2$\sigma$ contour is outlined. The blue and red stars in the corner-plot mark the best and median solutions. The same is indicated with the vertical blue line (best solution) and dotted/dashed red lines (median and 16-84\% percentiles). The mass-weighted age and the current stellar mass are reported.}
 \label{corner}
\end{figure*}

Based on the \JWST/NIRCam photometry presented in LB25 (from their Table~3) and adopting the new  spectroscopic redshift $z=9.625$, we perform here a new SED analysis of the CI with {\tt Bagpipes} \citep{carnall19}. Following the analysis of MM25 we use BPASS v2.2.1 templates \citep{Eldridge2017}, \citet{Kroupa2002} stellar initial mass function (IMF), delayed-$\tau$ star-formation history (SFH), \citet{Calzetti2000} dust attenuation curve. Driven by the new spectroscopic constraints and the discussion in MM25 we assume that the CI is not younger than the hosted clusters ($> 10$ Myr). We also explore the effect of the \lya-damping (Christensen et al. in prep.) on the inferred physical properties by including/excluding the F150W band from the fitting (the best-fit results do not change the conclusions of this work, see Appendix~\ref{SED_FIT}).  
The inferred fiducial physical quantities and uncertainties are reported in Figure~\ref{corner}. 

The intrinsic stellar mass (corrected for $\mu_{host}$) of the CG galaxy $-$ derived from the CI $-$ is $m_{CG} = 3.5\times 10^7$ 
\msun\ with 68\% central interval $(1.7 - 6.9)\times 10^{7}$ \msun\ and a mass-weighted age of 13 ($9 - 25$) Myr. Within the uncertainties, the stellar mass is in agreement with the values inferred from the arc $(2.5 - 5.6)\times 10^{7}$ \msun\ reported by LB25, which, however, assumed the previous $z=10.2$ photometric redshift. 
It is important to notice that the mass of the arc from LB25 has been corrected by a median magnification value, but the magnification gradients across the system are severe. Thus, the agreement 
between the two mass estimates (the arc and the CI) gives us confidence that the CG mass is close to this value. The age recovered for the CI agrees with the age ranges obtained for the star clusters (between 8 and 28 Myr, MM25).
Overall, the estimated formation age of the CG galaxy is likely not older than 30 Myr and mainly refers to the age of the recent burst of star-formation (see also discussion in Sect.~\ref{SFH} about uncertainties related to the SFH).

Figure~\ref{galfit} shows the {\tt Galfit} modeling \citep[][]{Peng_2010} of the CI detected at S/N~$\simeq 15$ in the NIRCam F150W band, the one with the sharpest PSF ($\simeq $ 0\farcs05). The galaxy appears resolved and nucleated with an effective radius R$_{\rm eff} = 1.6_{-0.2}^{+0.3}$ pixels (1 pixel = 0\farcs02) and magnitude $28.5\pm 0.2$ (in agreement with LB25). We fixed the Sersic index to $n = 1.0$, but notice that similar results are obtained adopting a Gaussian ($n=0.5$) profile. 
As reported in Figure~\ref{galfit}, the {\tt Galfit} modeling produces an excellent residual map (reduced $\chi^2 = 0.63$). The intrinsic effective radius R$_{\rm eff} = 1.6 [pix] \cdot 0.02'' \cdot 4360 [pc/''] / \mu_{host}^{-0.5} = 103_{-15}^{+13}$ parsec implies
$\Sigma_{\rm mass} = \frac{1}{2}~3.6 \cdot 10^{7} / (\pi\ \times$~R$_{\rm eff}^{2}) \simeq  520_{-225}^{+340}$~\msun~pc$^{-2}$ (or $5.3\times 10^{8}$~\msun~kpc$^{-2}$), where the errors account for R$_{\rm eff}$, $\mu_{host}$ and mass uncertainties. 
Similarly, adopting a constant star formation rate (SFR) over the last 20 Myr, the $\Sigma_{\rm SFR}$ is $17_{-7}^{+11}$ \msun~yr$^{-1}$~kpc$^{-2}$. 
The stellar mass and size of the CI is comparable with those inferred at similar redshift and luminosity \citep[][]{morishita2024_sizes, Ono2023, Tang2023}.
Taking into account the area underlined by a two-sigma contour in the F150W+F200W image (see Figure~\ref{corner}, corresponding to $0.141$~kpc$^{-2}$ in the source plane), the above quantities $\Sigma_{\rm SFR}$ and $\Sigma_{\rm mass}$ still remain quite large: 
$8$ \msun~yr$^{-1}$~kpc$^{-2}$ and $229$~\msun~pc$^{-2}$, respectively. Interestingly, the high stellar mass surface density ($\Sigma_{\rm mass}$) inferred here is comparable to that observed in compact galaxies at $z \gtrsim 10$, including the recently discovered $z = 14.44$ MoM-z14 \citep[][]{Naidu2025}, one of the most nitrogen-enhanced sources identified with JWST ($[{\rm N/C}] > 1$). In an emerging picture suggesting a size–chemistry bimodality at $z > 10$, where extended systems tend to be nitrogen-poor while compact galaxies exhibit strong nitrogen emission \citep[e.g.,][]{Naidu2025, Ji2025}, the CG galaxy appears to share the same stellar density and morphology as the nitrogen-rich class, reminiscent of similarly dense and massive star clusters discovered at lower redshift like the {\tt Sunburst} at z=2.37 \citep[][]{mestric2023, Schaerer2024}. 
It is also worth mentioning that the large sSFR of the CG galaxy experienced in the past (up to 100 Gyr$^{-1}$, see Sect.~\ref{SFH}) also favors an N-enhancement scenario \citep[][]{Topping2025}. It could likely be observed during its ``off-mode'' star-formation, when most of its stellar mass is already assembled in bound clusters, consistent with a globular-cluster-like environment.

The SFR inferred above for the CG galaxy under the assumption of constant star formation yields a lower limit on $\Sigma_{\rm SFR}$ that already exceeds values typical of local galaxies (at least under current constraints we have). This implies a phase shorter than $\sim$20 Myr during which $\Sigma_{\rm SFR}$ was higher than the estimate above, likely linked to efficient star-cluster formation.
This conclusion follows from comparisons with local star-forming galaxies showing $\log(\Sigma_{\rm SFR})>0$, which are generally associated with star-cluster formation efficiencies $\Gamma \gtrsim 50\%$ \citep{adamo20extreme}, measured over comparable timescales (10–20 Myr). However, it is important to stress that, in local galaxies, the mass in star clusters is insignificant with respect to the total mass of the galaxy.  This is not the case in galaxies like the CG.
It is worth noting that the formation of five massive star clusters in the last 30 Myr is remarkable in such low mass galaxy. As a reference, for typically observed conditions in the local Universe, the minimum total stellar mass a galaxy needs to form in order to sample one $10^6$~\msun\ cluster is $\simeq 5 \times 10^{7}$~\msun\ \citep[e.g.,][]{elmegreen12, elmegreen_Debra2017}, which is the same amount inferred in the CG galaxy.
The presence of five such massive clusters in the CG galaxy is statistically unexpected and therefore poses an interesting question on the possible large mass fraction in star clusters in this early galaxy. This is addressed in the next section.

\section{A preponderance of stellar clusters in the Cosmic Gems arc }
\label{SCMF}
The quantity $\Gamma$ reflects the clustering in space and time of the star formation process, which spreads the stellar mass into many individual clusters with a mass distribution often expressed as a power-law slope
of $-$2 (for linear intervals of mass) and an exponential drop at
some high cluster mass, M$_c$ \citep{elmegreen12, adamo2020SSRv}.
Depending on the environmental conditions, M$_c$ can be as high as $10^7$ \msun\ \citep{adamo20extreme} and the star cluster mass function (SCMF) is essentially a power law up to the most massive star cluster formed \citep{elmegreen_Debra2017}. In this analysis we test different assumptions for $\Gamma$ as described below.

For the sampling of the cluster mass function we adopt the functional form $dn(M)/dM~=~n_{0}M^{\beta}$, assume three slopes $\beta \simeq -1.5, -2.0$ and $-2.5$, and integrate down to 3 different (low) mass limits, M$_{\rm lim} = 10^2$, $10^{3.5}$, and $10^5$ \msun. 

Firstly, we determine here the fraction of stellar mass residing in star clusters with respect to the stellar mass of the CG galaxy, using the reported masses of the 5 star clusters from MM25. This estimate is obtained as follows: the sum of the mass in the five massive star clusters is used to normalize the mass distribution of the entire star cluster population, which is integrated down to a given m$_{\rm lim}$ \citep[][]{elmegreen_Debra2017}.  
In particular, considering the best delensed estimates of the stellar masses of the five star clusters (6.02, 2.24, 1.05, 0.78, 0.93$\times 10^{6}$ \msun\ for clusters A, B, C, D and E, respectively indicated in Figure~\ref{galfit}) 
and integrating the SCMF down to m$_{\rm lim}=10^2$ \msun\ with $\beta=-2$, we obtain a total mass in clusters ($5.9 \times 10^7$ \msun) which is 1.68 times higher than the intrinsic mass of the CI ($3.5 \times 10^7$ \msun).
This simple calculation already suggests that the CG galaxy formed a population of star clusters $-$ including massive ones $-$ very efficiently, in a way that must be consistent with the total stellar mass of the burst that produced them. However, the calculation must incorporate the uncertainties in the observed cluster masses. We account for these uncertainties by propagating them, via Monte Carlo (MC) sampling, as described below.

\subsection{The fraction of star cluster mass in the Cosmic Gems galaxy}

\begin{figure*}
\center
 \includegraphics[width=\textwidth]{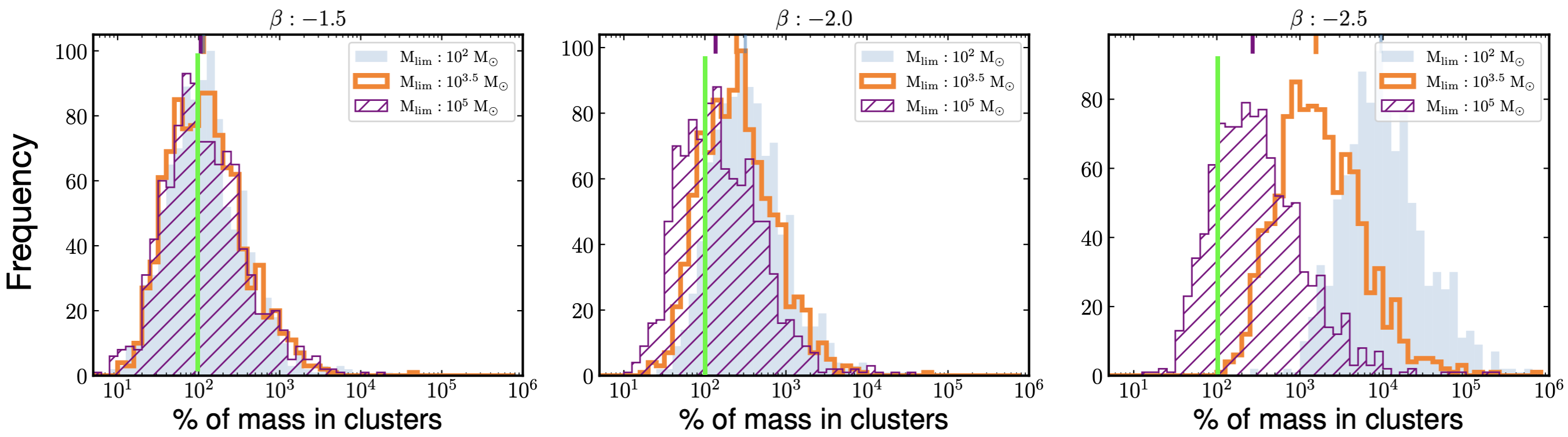}
 \caption{
 Monte Carlo realizations of the fraction of the stellar mass of the CG galaxy residing in the population of bound star clusters is shown (uncertainties on the normalization of the SCMF and the stellar mass of the host galaxy are included, see text for details). From left to right the SCMF is evaluated for slopes $\beta$ of $-1.5$, $-2.0$, and $-2.5$, adopting three different low mass limits as indicated in the legend of each panel. Calculations have been performed assuming the fiducial magnification values (while the behavior with varying magnification is shown in Figure~\ref{multi_hist}). The vertical line bar marks the case where the total stellar mass of the star clusters equals that of the host galaxy (fraction equal to 100\%).
 }
 \label{distrib}
\end{figure*}
\begin{figure}
\center
\includegraphics[width=\columnwidth]{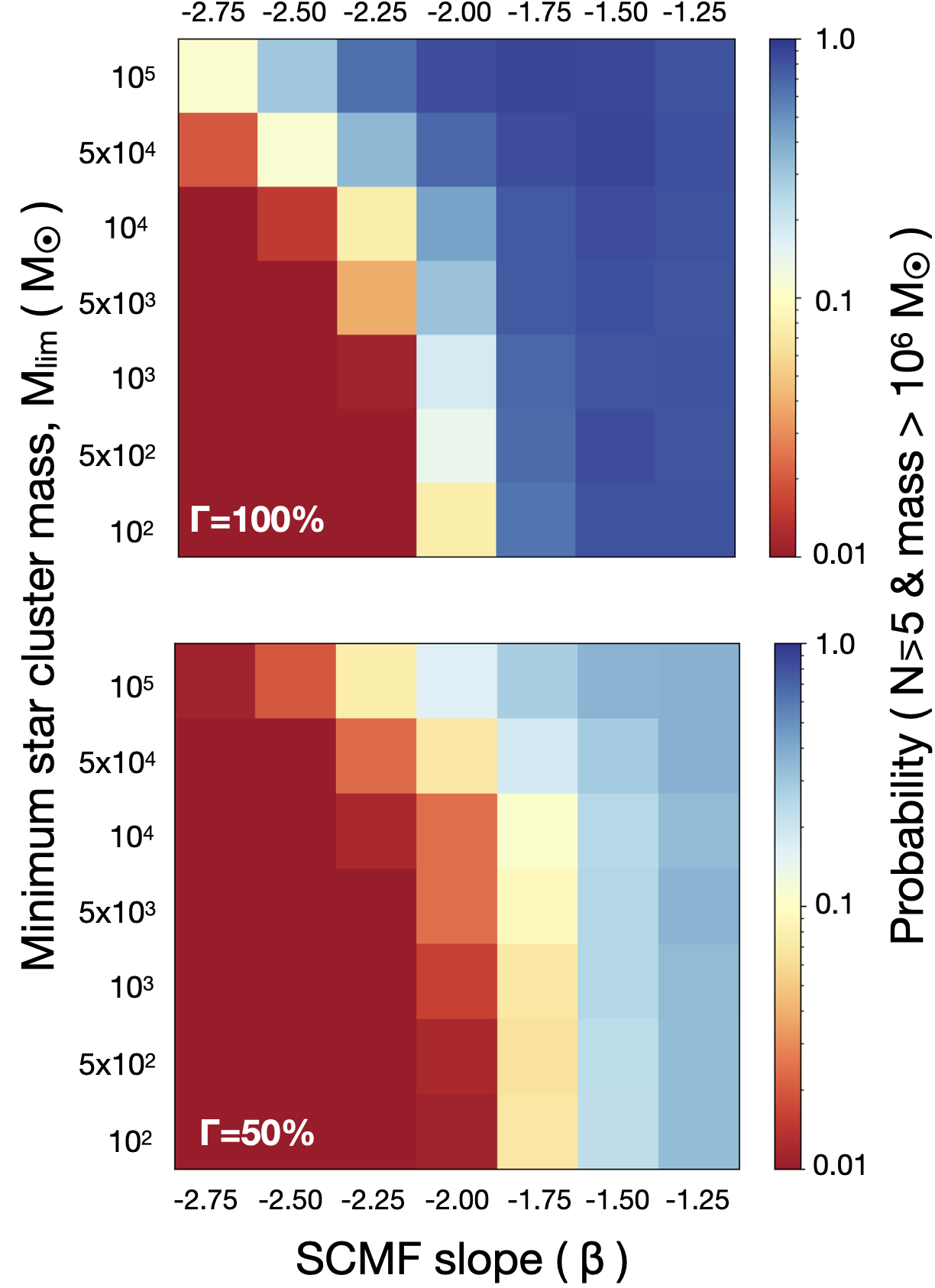}
 \caption{Statistical sampling of the SCMF. The color-coded probability of having 5 or more massive star clusters (with masses exceeding $10^6$~\msun) is shown as a function of the slope of the stellar cluster mass function (SCMF, $\beta$) and the low mass limit used to integrate the SCMF. The top panel represents the scenario where the entire mass of the CG galaxy is composed of stellar clusters ($\Gamma = 100\%$), while the bottom panel illustrates the case with $\Gamma=50\%$.
 }
 \label{probes}
\end{figure}

The normalization of the SCMF depends on the lens model magnification of each star cluster.
In the following we report the results adopting the updated fiducial values of magnification for each star cluster and $\mu_{host} = 1.84 \pm 0.05$ (the effect of different magnifications is shown in the Appendix~\ref{mass_mu}). 
We include the uncertainty of the stellar masses through a MC process, which extracts 1000 realizations of the five star cluster masses drawn from distributions following their uncertainties. For each set of masses (realization) we have the minimum (m$_{\rm min}$), maximum (m$_{\rm max}$) and the sum of masses of the actual set of five clusters (S$_{\rm clusters}$). The normalization of the SCMF is properly calculated at each MC realization by requiring the integrated portion of the SCMF between m$_{\rm min}$ and m$_{\rm max}$ is S$_{\rm clusters}$ (the same results are obtained if we consider the number of clusters in place of integrated mass).

Once the normalization is calculated at the given MC extraction, the inferred stellar mass of the full star cluster population ($m_{\rm SC}^{\rm tot}$) integrated in the mass range M$_{\rm lim}$ to m$_{\rm max}$ is then compared to the delensed stellar mass of the CG galaxy, $m_{\rm CG}$ (inferred from the CI, including its uncertainties on the mass and $\mu_{host}$). 
M$_{\rm lim}$ is the adopted minimum cluster stellar mass. 
The resulting mass fraction ($m_{\rm SC}^{tot}$/m$_{\rm CG}$) of the stellar mass located in the star cluster population calculated adopting the fiducial magnification case and varying the slope of the mass function $\beta=-1.5, -2.0, -2.5$ and the M$_{\rm lim}$ is shown in Figure~\ref{distrib} (the behavior as a function of the magnification is discussed in Appendix~\ref{mass_mu}). Depending on the assumptions, the distributions in the figure show that the stellar mass of the CG galaxy cannot indiscriminately accommodate all the solutions.  
The galaxy does not have enough stellar mass to accommodate a cluster population with a SCMF of slopes $-$2.5 even when a very high M$_{\rm lim}$ is used  (nearly all realizations exceed the mass of the host galaxy). On the other hand, a top-heavy SCMF with a slope of $-2.0 < \beta < -1.5$ and increasingly high M$_{\rm lim}$ produces mass fractions peaked at 100\%, i.e., it is more likely that the galaxy has sufficient mass to accommodate a cluster population, even if there is still a large fraction of the host mass located in star clusters. 
In general, if the SCMF is sampled down to a mass limit of $10^4$~\msun\ or lower, it is more likely that it is top-heavy.  If we look at the mass fraction as a measurement of $\Gamma$, this exercise implies that cluster formation efficiency is very high and almost reaching unity: nearly the full mass of the galaxy is located in star clusters. 
While it has been argued that the stellar initial mass function (IMF) may turn top-heavy at high redshift and/or low metallicity \citep[e.g.,][]{Chon21,Steinhardt23, Meena2025}, adopting a top-heavy IMF for the CG galaxy is unlikely to challenge these conclusions. If the actual IMF indeed features a flatter slope than the standard IMF, or an extension to much higher stellar masses, then an analysis based on the standard IMF could cause the total stellar masses of young stellar systems to be overestimated. However, since the CG star clusters and the CI display similar SED shapes (and consequently similar estimated ages), such putative mass offsets are likely to be similar in the star clusters and the overall galaxy, thereby canceling any significant effect on the inferred mass fraction of clusters.

\subsection{Stochastic Sampling of the star cluster mass function}

In this section, we derive the likelihood of forming at least five massive star clusters with masses $> 10^6$ \msun\ assuming a fraction of the galaxy mass $m_{\rm SC}^{tot} = m_{CG} \cdot \Gamma$ is in bound star clusters. We adopt $m_{CG} = 3.5\times 10^7$ \msun\ (the stellar mass of the host galaxy, Sect.~\ref{physical_properties}) and $\Gamma = 1.0$ (100\%, the full galaxy mass) and $\Gamma = 0.5$ (50\%, half of the galaxy mass). The random sampling of the SCMF is performed assuming slopes $\beta$ from $-$2.5 to $-$1.5 (with step 0.25), low mass limit (M$_{\rm lim}$) from $10^2$ \msun\ to $10^5$~\msun. The upper mass end of the distribution, i.e., the maximum cluster mass, is chosen as half of the available galaxy mass in clusters ($0.5 \cdot m_{CG} \cdot\Gamma$). One thousand realizations have been performed for each combination of parameters. The process stochastically generates for each realization a synthetic population of star clusters that obeys  the aforementioned parameters set. In particular the specific masses of individual clusters and the number of clusters vary from run to run (since high-mass clusters are rare, they ``use up'' more of the total budget), the presence or absence of massive outliers depends on chance (especially near the maximum cluster mass).

The resulting probabilities of forming at least five massive star clusters (mass $> 10^6$~\msun) are shown in Figure~\ref{probes}. Consistent with the results discussed in the previous section, a top-heavy SCMF with slope shallower than $-2$ ($-1.8, -1.5$) and/or higher minimum cluster mass limit (M$_{\rm lim}$) is clearly preferred. Only by maximizing $\Gamma$ to 100\% and pushing the low mass limit of the mass function to higher values, does it allow for a solution with slopes close to $-2$. In general, solutions with slopes steeper than $-2$ have very low ($\lesssim1$\%) likelihood under all assumptions. 

These results are, to some extent, dependent on the effective resolution and our ability to identify individual massive clusters. 
Although it is unlikely that parsec-scale objects are composed of unresolved sub-components (i.e., lower-mass star clusters), it is worth noting that the same result is obtained even when the cluster masses are halved and their number doubled. Finally, shallow slopes of the SCMF ($-1.5$ or $-1.25$) might overproduce the number of massive star clusters relative to the observed population (i.e., those with masses above $10^6$~\msun) and rapidly saturate the fraction of the CG stellar mass assumed to reside in star clusters. This results in an overall lower probability of finding at least five massive clusters at small values of $\Gamma$, as illustrated in Figure~\ref{probes} by comparing the top and bottom panels. Even in the case of $\Gamma=100$\% (i.e., $3.5\times10^7$ \msun\ resides in star clusters) $-$ the most favorable scenario for forming massive clusters $-$ the expected number of massive clusters for a top-heavy SCMF rarely exceeds 10, with the probability of having at least 10 clusters more massive than $10^6$~\msun\ remaining below 5\%.
Finally, we verified that the results remain unchanged when adopting a Schechter formalism with $M^\star$ equal to the cluster-mass cutoff described above, $0.5 \cdot m_{CG} \cdot \Gamma$, and the same slope on the power-law side.

The observed massive star clusters in the CG arc can be attributed to a combination of factors, including a top-heavy star cluster mass function (SCMF), combined with a high fraction of the 
host galaxy's star formation occurring in bound star clusters (high $\Gamma$), and/or the suppression of star cluster formation at the low-mass end. Various feedback mechanisms, 
such as radiation, supernovae (SN), winds and Lyman-alpha (Ly$\alpha$) pressure, influence the initial shape of the SCMF \citep{Andersson2024_preSN, Nebrin2025}, which is ultimately connected to the IMF within clusters \citep{Elmegreen2006_IMFstellar_cluster, krumholz2019}. 
These physical mechanisms can affect the shape of the SCMF, but high-resolution simulations of low-metallicity star cluster formation that include this physics are, for the most part, still missing.

Using a cosmological zoom simulation, \cite{Sugimura2024} found that the formation of massive bound star clusters and the extreme burstiness observed in several high-z galaxies is mainly produced by the effect of strong Lyman-Warner radiation (FUV) from Pop~II stars in a low-metallicity environment. This radiation leads to a hot ISM through suppression of H$_2$ formation and cooling, thus increasing the Jeans mass and the typical masses of star-forming clouds. As shown in \cite{Garcia2023, Garcia2025}, this mechanism leads to a flattening of the SCMF rather than an increase of the low-mass cutoff of the mass function. The properties of the compact star clusters in these simulations are in good agreement with those found in the CG, however, there are some caveats. The galaxies simulated by \cite{Garcia2025} and \cite{Sugimura2024} have relatively low mass and low metallicity with respect to the CG \citep{Messa_D1T1_2025}. In more massive galaxies, invoking simple scaling arguments, we may expect stronger bursts of star formation and more massive gas clouds for a given power-law slope of the SCMF. These scaling relations may balance out the effect of increased gas metallicity, but more simulations of higher-mass galaxies need to be run to confirm this hypothesis. 
In addition, it is also possible that some physical ingredients are still missing in these simulations that may further increase the Jeans mass in the ISM even at higher metallicities, for instance heating from Cosmic rays and X-rays from binary stars, accreting IMBHs or micro-AGN.

The conditions for star formation in these high-redshift compact galaxies (the 5 clusters are located in a region spanning about 70 pc, while from the CI we derive a very small size for the entire galaxy) drive very high star formation surface densities which, in turn, contributes to elevated $\Gamma$ values \citep{adamo20extreme, CFE_li18}. Recently, Ly$\alpha$ feedback has been proposed as a dominant process that might suppress the formation of low-mass star clusters \citep{Nebrin2025}.
Despite the very high lensing magnification factors, completeness limitations related to resolution and depth hinder the ability to count or disentangle individual star clusters down to low masses ($10^3 - 10^4$ \msun). The lack of detection of low mass clusters in the CG arc might imply that they have been suppressed, however we cannot exclude incompleteness. 
These limitations will be addressed in a forthcoming study that will employ forward modeling simulations incorporating lensing uncertainties.
Assuming a slope $\beta= -2$, in line with the observed local values for young starbursts ($\lesssim 50$ Myr; \citealp[e.g.,][]{whitmore2010, linden2021,adamo20extreme}), and with those derived from simulations \citep{Calura2025, Pascale25, he20_ricotti, krumholz2019}, our results suggest that large $\Gamma$ values are preferred in the CG galaxy along with a higher low mass limit.
This study is the first attempt to investigate the properties of the star cluster mass function and formation efficiency at 
such small spatial scales in the early Universe (first half Gyr after the Big-Bang). The results are based on observations of a single galaxy, and additional 
statistical samples will be necessary to confirm whether the 
occurrence of massive star clusters in relatively low-mass 
galaxies is a common phenomenon at early epochs.

\section{Star-formation history}
\label{SFH}
The observed magnitude of the CI in the F200W filter is $28.4\pm0.1$ (LB25), corresponding to an intrinsic magnitude of $29.1\pm0.1$ after accounting for the lensing magnification factor, $\mu=1.84$. This translates to an absolute UV magnitude of M$_{\rm UV}^{obs} = -18.4 \pm 0.1$ at rest-frame $\lambda \simeq 1800$ \AA.
This value is about 4 times fainter than the so-called ``blue monster'' regime of the recently identified class of bright galaxies at $z>9$ showing M$_{\rm UV} \lesssim -20$ \citep[][]{Whitler2025, Napolitano2024, Finkelstein2024, McLeod2024, Castellano2023, Donnan2024, Harikane2023, Harikane2024, Perez-Gonzalez2023, Tang2025, Donnan2025}.

Could the CG galaxy have experienced such a phase in the past?
The typical and fundamental source of uncertainty on the SFH arises from the spatially unresolved formation histories of the internal regions, which contribute collectively to the observed integrated light. In addition, the absence of rest-frame optical coverage also prevents us from deriving any solid conclusion about the presence of old ($\gtrsim 100$~Myr, or $z>11.4$ in the present case) stellar populations, and limits us to infer an ultraviolet/B-band-based SFH. This limitation also applies to the CI discussed here. However, the strongly magnified arc offers critical insight on the formation histories of individual sub-components of the galaxy.
The ages of the individual star clusters span a time interval consistent with the formation time-scale inferred from the CI ($9 - 27$ Myr, Figure~\ref{corner}). In particular, four out of five clusters formed between 8 and 15 Myr ago (A, B, C, D), while the fifth cluster (E) dates back to approximately 27 Myr ago (MM25); such a post-burst mode of the CG galaxy is also corroborated by weak optical emission lines observed in new spectroscopic \JWST/NIRSpec observations (MM25). With a total intrinsic stellar mass of $\simeq 1.1\times10^7$ \msun\ located in clusters A$-$E, formed during an interval of time $\sim 20$~Myr, 
the resulting minimum sSFR  is sSFR~$\simeq 50$~Gyr$^{-1}$ (adopting constant SFR). Assuming the last burst made the bulk of the mass that we inferred from SED fitting ($3.5\times10^{7}$ \msun, see Figure~\ref{corner}), 
then the CG galaxy was very active in forming stars in the recent past ($\gtrsim 10$ Myr ago) and likely appeared brighter in the ultraviolet than observed now.

\begin{figure}
\center
\includegraphics[width=\columnwidth]{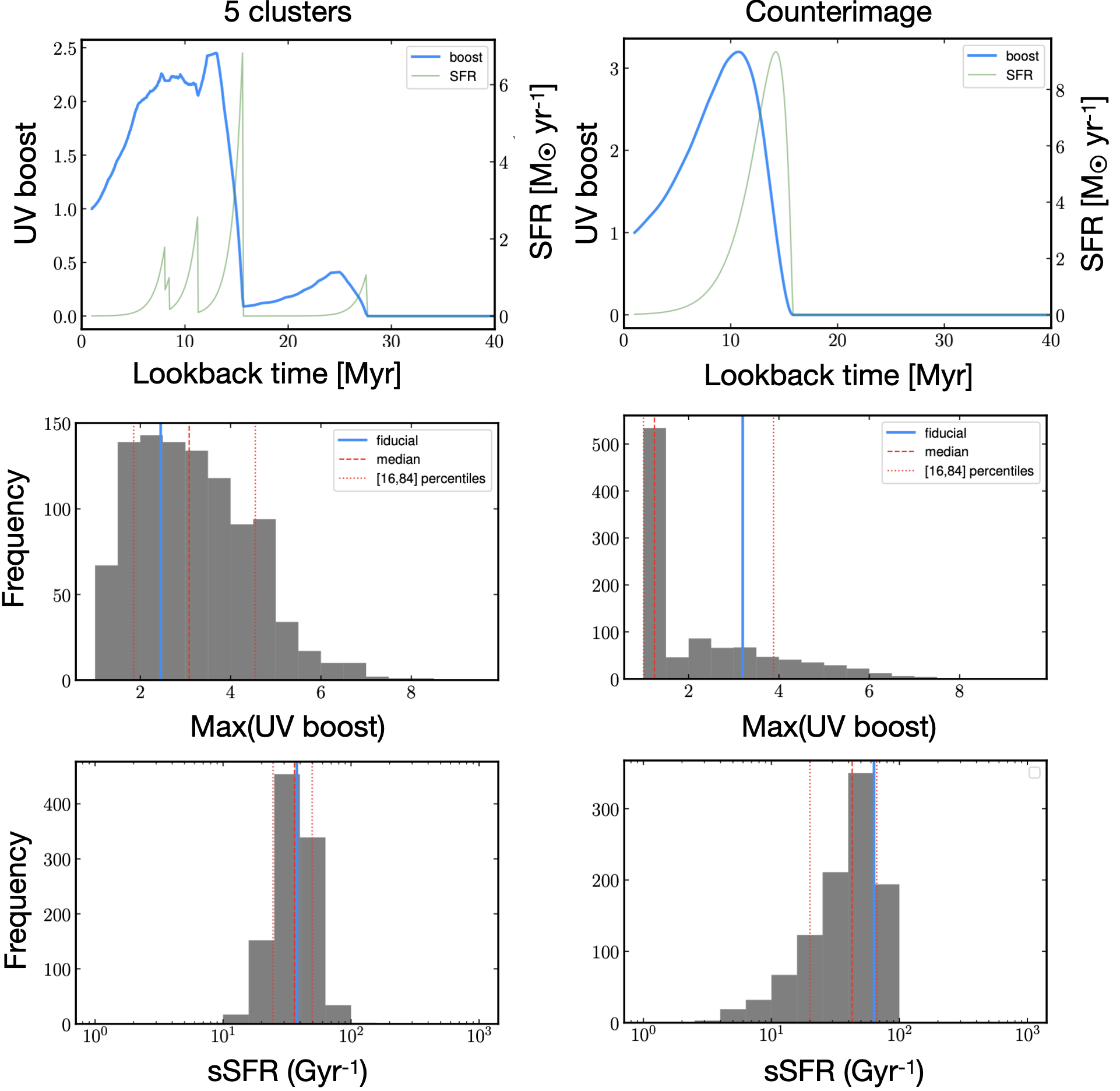}
 \caption{Inferred past star formation activity for the five star clusters (left panels) and for the CI (right panels). From top to bottom: the best-fit solutions of the UV boosting factor (blue curve) and SFR (green curve); Monte Carlo distribution of the maximum boosting factor with median, percentiles and  fiducial values, see text for details; Monte Carlo distribution of the sSFR (median, percentiles and fiducial values follow the same color-coding as in the middle panels).}
 \label{CI_SFH}
\end{figure}

This aligns with the relatively short formation timescale inferred from the SED fitting, which describes the dominant UV-weighted mass assembly of the CG galaxy. Figure~\ref{CI_SFH} (top panels) shows the fiducial SFHs obtained by the SED fit for the CI (left), and for the sum of each individual star cluster (right). 

Rather than the CI, we focus on the star clusters, which more directly trace the most recent burst of star formation.

Two main effects drive the recent luminosity history of the CG galaxy: (1) stellar aging and (2) dust extinction.
To assess the ultraviolet luminosity of the CG, we trace back the SFH of each cluster, using BPASS models to determine its age and mass at each timestep\footnote{all other quantities, i.e., extinction, metallicity and logU, are left constant with time.}, and consequently its UV luminosity. The sum of these luminosities, compared to the observed one, gives the UV boosting factor (Figure~\ref{CI_SFH}, top). In order to assess the uncertainties, we run 1000 Monte Carlo realizations; in each realization we randomly draw the properties of each star cluster (most relevantly, age and mass) from the posterior distribution used, in the first place, to derive its best-fit properties in MM25 (see their Table~2). We then repeat the process for finding the evolution of the ultraviolet boosting factors, and in particular of the maximum boost reached in each realization, whose distribution is shown Figure~\ref{CI_SFH} (central). The median maximum boosting factor is 3.1 with 16th/84th percentiles 1.9/4.5.
In magnitudes, the median brightening from aging alone is therefore $-1.23$, occurring $\simeq 15$ Myr ago.

The current photometric $A_V$ estimates for the five clusters span $\simeq 0.2 - 0.4$, while the host galaxy has $A_V\simeq 0.23$ (MM25), within 1$\sigma$ uncertainty of $A_V$ inferred from the CI. Adopting $A_V\simeq 0.2$, the corresponding $A_{1800}$ is $\simeq 0.5$ mag, computed from $A_{1800}=A_V k(1800)/R_v$ using the Calzetti law ($k(1800)\approx 9.4$, $R_V=4.05$). Assuming the same attenuation for the host galaxy, the dust-free M$_{\rm UV}$ of the CI is $-18.9$ (from the observed $-18.4$ corrected by 0.5 mag). Therefore, in the absence of dust attenuation and accounting for aging alone, the CI was intrinsically brighter than M$_{\rm UV}=-20$ in 57\% of the MC realizations. 

The key question is whether there existed a phase in which dust attenuation decreased sufficiently to resemble a ``blue monster'' episode (i.e., high sSFR, very blue UV slope, UV-bright). The time evolution of $A_{1800}$ is not directly measurable and must be modeled. 

The sSFR of the system is measured as the inverse of the timescale needed to form its total mass\footnote{In the 5-cluster case, this time is set by the onset of the SFR event for the oldest cluster.}; the values found are conservative estimates for the average sSFR during the entire SFR event, but larger values could have been reached in the case of a bursty SFH. Also in this case, the posterior distributions from the SED fit are used to estimate uncertainties (Figure~\ref{CI_SFH}, bottom panels). 
The UV magnitude of the CG galaxy might have experienced significant enhancement in the past, associated with periods of peaked sSFR. 
The data suggest that in more than 55\% of the cases the sSFR exceeded $25$~Gyr$^{-1}$ during the last burst, with values reaching 100 Gyr$^{-1}$ (Figure~\ref{CI_SFH}). 
This value meets the critical threshold of 25 Gyr$^{-1}$ proposed by \citeauthor{Fiore2023} (\citeyear{Fiore2023}; \citealp[see also][]{Ferrara_2023_monsters})
as the condition for the onset of radiation-driven outflows, which has been suggested (among other mechanisms; e.g., \citealt{Finkelstein2024}) to explain the overabundance of bright blue galaxies observed at $z>9$.
In particular, in this scenario outflows develop when a galaxy experiences a super-Eddington phase boosted by stellar radiation in compact and dusty galaxies. Recently, \citet[][]{Nakazato2024} studied 20 galaxies at $z>10$ and investigated if they experienced a dusty outflow phase in their recent past. The same analysis of \citet[][]{Nakazato2024} applied to the CG galaxy suggests that it was indeed able to develop such a radiation driven outflow about 10~Myr before observations, consistent with the above mentioned sSFR activity and the recent burst traced by the ages of clusters A, B, C, D about $8-15$ Myr ago. Based on these tests, the CG galaxy would have appeared as a ``Blue Monster'' had it been observed a few million years earlier, during its bright phase.

The more time-compressed the star-cluster formation events are, the greater the resulting boost in both ultraviolet luminosity (modulo dust attenuation) and the galaxy-wide specific star-formation rate (sSFR). At very high redshift, the stochastic nature of star formation further amplifies this UV enhancement. However, this effect alone still appears insufficient to account for the observed overabundance of bright $z\sim10$ galaxies \citep[e.g.][]{Pallottini2023, Carvajal_Bohorquez2025}.

The high cluster-formation efficiency, $\Gamma$, measured for the CG galaxy (Section~\ref{SCMF}), combined with its low dust attenuation, also supports the feedback-free scenario proposed by \citet{Dekel23} (see also \citealt{Williams2025_LambdaCDM}). In this framework, rapid gas accretion is converted into stars with nearly unity efficiency, triggering successive starbursts that enhance the ultraviolet luminosity. The bound stellar remnants from these bursts subsequently appear as massive star clusters. 

Ultimately, a combination of the aforementioned mechanisms is likely responsible for shaping the $z>9$ ultraviolet luminosity function. Additional spectroscopic observations of CG-like systems are needed to further investigate these scenarios.

\section{Final remarks}

The Cosmic Gems galaxy is the first example of a $z\simeq 10$ source in which parsec-scale star-formation and individual stellar clusters are probed. 
We observe five massive star clusters likely formed within the last 30 Myr.
We assume that the star cluster mass function (SCMF) follows a power-law distribution with a given slope. After normalizing the high mass-end to the five massive clusters, and assuming a high-mass cutoff (taken as half of the galaxy’s mass), a minimum star cluster mass in the integration (M$_{\rm lim}$), and a value for the cluster formation efficiency ($\Gamma$), the total stellar mass in clusters exceeds 
the stellar mass formed during the host galaxy's most recent burst in 
several combinations of explored SCMF slopes and M$_{\rm lim}$.
The results of this exercise suggest that for $\Gamma=0.5$ (and even $\Gamma=1$), a fully populated SCMF with slope $\beta=-2$ or steeper and a lower mass cutoff M$_{\rm lim}<10^4$~M$_\odot$ is highly unlikely, as it would predict more mass than is actually available from the galaxy’s last burst. 
Solutions that favor larger values of M$_{\rm lim}$ may point to two, not mutually exclusive, effects:
(i) the formation of low mass bound star clusters might be inhibited due to early feedback \citep[e.g., \lya\ feedback,][]{Nebrin2025}; and/or
(ii) an environmental dependence related to the ISM surface density, such that the higher the density the higher the minimum mass of bound stellar clusters that reshapes the initial SCMF \citep[e.g.,][]{Trujillo-Gomez19}.

Although it remains to be tested whether, by $z\simeq10$, sufficient time has elapsed for low-mass clusters to be destroyed by intense stellar feedback, tidal shocks, and other dynamical processes, it is also noteworthy that disrupted clusters would significantly contribute to the total stellar mass of the galaxy. Flatter SCMF slopes than the canonical $\beta=-2$ have also been reported in cosmological simulations of galaxies at $z>8$. \citet{Garcia2023} found slopes ranging from $\beta\sim-1.4$ to $-1.7$, with a tendency for the slopes to steepen over time following each starburst episode. After the initial burst of star formation at $z>10$, they measured $\Gamma$ values between 0.5 and 1. More recently, \citet{Garcia2025}, using an improved sub-grid prescription incorporating a physically motivated star formation efficiency in star-forming gas clouds, found even flatter slopes of $\beta\sim-1.3$, in agreement with the findings presented in this paper.

In summary, by analyzing the Cosmic Gems' arc and the counter-image, we find the following results:

\noindent $\bullet$ 
Under reasonable assumptions about the shape of the SCMF, normalized to the observed high-mass end, we find that the CG galaxy's stellar mass occurred during the last burst is not high enough to fit a fully populated SCMF with slope $-2$, integrated down to a low cluster mass limit of M$_{\rm lim}=10^2$ \msun\ implying unrealistically high $\Gamma >> 100$\%. 
To reconcile the total stellar mass in cluster with the stellar mass of the burst in the host galaxy, a top-heavy SCMF with $\beta > -2$ and/or a high M$_{\rm lim}$ are required, along with a large fraction ($\Gamma = 50 - 100\%$) of the galaxy mass located in star clusters. These results weakly depend on the assumed magnification of the arc. 

\noindent $\bullet$
The CI provides a comprehensive view of the CG galaxy. Based on SED fitting and {\tt Galfit} modeling, the stellar mass and effective radius are estimated to be $3.5 (1.7-6.9)\times10^{7}$ \msun\ and $\simeq 100$ pc, respectively, which implies a high stellar surface density of $\Sigma_{mass} \simeq 520$ \msun~pc$^{-2}$.
The currently delensed ultraviolet luminosity (M$_{\rm UV}=-18.4$) and the presence of massive star clusters spanning the age interval $8-27$~Myr in which the bulk of hot and massive stars already died (MM25), suggest that this galaxy was more luminous in the past, potentially encompassing the ``Blue Monster'' regime. 
The detailed star formation history is currently limited by the \JWST/NIRCam photometry and NIRSpec spectroscopy probing ultraviolet/B-band rest-frame wavelengths. 
However, it is worth noting that the more compressed the star cluster formation events are (back in time), the higher the intrinsic luminosity of the galaxy. The best-fit solution from the SED-fitting and spectral analysis by MM25 suggests that the CG galaxy approached M$_{\rm UV} \simeq -20$ and likely experienced a large sSFR ($> 50$ Gyr$^{-1}$).

This is the first evidence that baryon concentration in the early Universe \citep[e.g.,][]{Renzini2025} was highly efficient in forming massive star-cluster-dominated systems, which likely played a pivotal role in driving the ionizing properties of early galaxies, as key agents of the reionization process \citep[][]{he20_ricotti}. Massive star clusters host (very) massive stars, eventually enhancing both the ionizing photon production efficiency \citep[e.g.,][]{Schaerer2025} and likely the escape fraction of ionizing photons \citep[e.g.,][]{vanz_sunburst, vanz22_CFE, rivera19, Rivera2024}.

Future large telescopes working with diffraction limited PSFs of $\sim$10 milli-arcsecond (e.g. ELT/MORFEO-MICADO, \citealt{MORFEO_2024, MICADO_2024_status}) will allow us to make a significant quantum leap when targeting moderately lensed galaxies, allowing us to probe parsec-scale physical regions ($<10$ pc) with modest magnification ($\mu < 10$). In fact, a magnification $\mu > \times 5$ is formally sufficient to reach the stellar cluster size regime (see discussion in \citet{vanz_mdlf} on future extreme adaptive optics facilities). 
Ongoing key \JWST/ massive surveys on lensed fields (e.g., the Vast Exploration for Nascent, Unexplored Sources (VENUS) large program with 300 hours allocated, n.6882 cycle 4, PI Fujimoto) will provide ideal targets for the extremely large telescopes. 
Observations of the CG arc with a 10 milli-arcsecond PSF resolution will enable sub-parsec light profile analyses at redshift $z\simeq 10$, along with potentially locating in the CI the region hosting the massive star clusters observed in the lensed arc.  

\begin{acknowledgements}
We thank the anonymous referee for helpful comments that improved the manuscript. EV thanks A. Ferrara for very useful discussions on the `blue monster' galaxy population. This work is based on observations made with the NASA/ESA/CSA \textit{James Webb Space Telescope} and \textit{Hubble} Space Telescope (\HST). 
The data were obtained from the Mikulski Archive for Space Telescopes at the Space Telescope Science Institute, which is operated by the Association of Universities for Research in Astronomy, Inc., under NASA contract NAS 5-03127 for JWST. These observations are associated with program \#4212 (PI L. Bradley) and \#5917 (PI E. Vanzella).
EV and MM acknowledge financial support through grants PRIN-MIUR 2020SKSTHZ, the INAF GO Grant 2022 ``The revolution is around the corner: JWST will probe globular cluster precursors and Population III stellar clusters at cosmic dawn,'' and INAF GO Grant 2024 ``Mapping Star Cluster Feedback in a Galaxy 450 Myr after the Big Bang'', and by the European Union – NextGenerationEU within PRIN 2022 project n.20229YBSAN - Globular clusters in cosmological simulations and lensed fields: from their birth to the present epoch. AA acknowledges support by the Swedish research council Vetenskapsr{\aa}det (VR 2021-05559, and VR consolidator grant 2024-02061). EZ acknowledges project grant 2022-03804 from the Swedish Research Council. This work was supported by JSPS KAKENHI Grant Numbers JP25H00662, JP22K21349 and JP23H00131.
J.M.D. acknowledges support from project PID2022-138896NB-C51 (MCIU/AEI/MINECO/FEDER, UE) Ministerio de Ciencia, Investigaci\'on y Universidades.
F.E.B acknowledges support from ANID-Chile BASAL CATA FB210003, FONDECYT Regular 1241005,
and Millennium Science Initiative, AIM23-0001.
AZ acknowledges support by grant 2020750 from the United States-Israel Binational Science Foundation (BSF) and grant 2109066 from the United States National Science Foundation (NSF), and by the Israel Science Foundation Grant No. 864/23.
Y.J-T. acknowledges financial support from the State Agency for Research of the Spanish MCIU through Center of Excellence Severo Ochoa award to the Instituto de Astrofísica de Andalucía CEX2021-001131-S funded by MCIN/AEI/10.13039/501100011033, and from the grant PID2022-136598NB-C32 Estallidos and project ref. AST22-00001-Subp-15 funded by the EU-NextGenerationEU.
Additionally, this work made use of the following open-source packages for Python, and we are thankful to the developers of: Matplotlib \citep{matplotlib2007}, MPDAF \citep{MPDAF2019}, Numpy \citep[][]{NUMPY2011}, QFitsView DPUSER language (https://www.mpe.mpg.de/~ott/dpuser/index.html).
\end{acknowledgements}

%
%

\bibliographystyle{aa}
\bibliography{bib}

@ARTICLE{Topping2025,
       author = {{Topping}, Michael W. and {Stark}, Daniel P. and {Senchyna}, Peter and {Chen}, Zuyi and {Zitrin}, Adi and {Endsley}, Ryan and {Charlot}, St{\'e}phane and {Furtak}, Lukas J. and {Maseda}, Michael V. and {Plat}, Adele and {Smit}, Renske and {Mainali}, Ramesh and {Chevallard}, Jacopo and {Molyneux}, Stephen and {Rigby}, Jane R.},
        title = "{Deep Rest-UV JWST/NIRSpec Spectroscopy of Early Galaxies: The Demographics of C IV and N-emitters in the Reionization Era}",
      journal = {\apj},
         year = 2025,
        month = feb,
       volume = {980},
       number = {2},
          eid = {225},
        pages = {225},
          doi = {10.3847/1538-4357/ada95c},
archivePrefix = {arXiv},
       eprint = {2407.19009},
 primaryClass = {astro-ph.GA},
       adsurl = {https://ui.adsabs.harvard.edu/abs/2025ApJ...980..225T}
}

@ARTICLE{Schaerer2024,
       author = {{Schaerer}, D. and {Marques-Chaves}, R. and {Xiao}, M. and {Korber}, D.},
        title = "{Discovery of a new N-emitter in the epoch of reionization}",
      journal = {\aap},
         year = 2024,
        month = jul,
       volume = {687},
          eid = {L11},
        pages = {L11},
          doi = {10.1051/0004-6361/202450721},
archivePrefix = {arXiv},
       eprint = {2406.08408},
 primaryClass = {astro-ph.GA},
       adsurl = {https://ui.adsabs.harvard.edu/abs/2024A&A...687L..11S}
}

@ARTICLE{Williams2025_LambdaCDM,
       author = {{Williams}, Claire E. and {Naoz}, Smadar and {Lake}, William and {Burkhart}, Blakesley and {Marinacci}, Federico and {Vogelsberger}, Mark and {Yoshida}, Naoki and {Menon}, Shyam H. and {Chen}, Avi and {Adamo}, Angela},
        title = "{{\ensuremath{\Lambda}}CDM Star Clusters at Cosmic Dawn: Stellar Densities, Environment, and Equilibrium}",
      journal = {\apj},
         year = 2025,
        month = sep,
       volume = {990},
       number = {2},
          eid = {135},
        pages = {135},
          doi = {10.3847/1538-4357/adf19d},
archivePrefix = {arXiv},
       eprint = {2502.17561},
 primaryClass = {astro-ph.GA},
       adsurl = {https://ui.adsabs.harvard.edu/abs/2025ApJ...990..135W}
}

@ARTICLE{Schaerer2025,
       author = {{Schaerer}, D. and {Guibert}, J. and {Marques-Chaves}, R. and {Martins}, F.},
        title = "{Observable and ionizing properties of star-forming galaxies with very massive stars and different initial mass functions}",
      journal = {\aap},
         year = 2025,
        month = jan,
       volume = {693},
          eid = {A271},
        pages = {A271},
          doi = {10.1051/0004-6361/202451454},
archivePrefix = {arXiv},
       eprint = {2407.12122},
 primaryClass = {astro-ph.GA},
       adsurl = {https://ui.adsabs.harvard.edu/abs/2025A&A...693A.271S}
}

@ARTICLE{Carvajal_Bohorquez2025,
       author = {{Carvajal-Bohorquez}, C. and {Ciesla}, L. and {Laporte}, N. and {Boquien}, M. and {Buat}, V. and {Ilbert}, O. and {Aufort}, G. and {Shuntov}, M. and {Witten}, C. and {Oesch}, P.~A. and {Covelo-Paz}, A.},
        title = "{Stochastic star formation activity of galaxies within the first billion years probed by JWST}",
      journal = {arXiv e-prints},
         year = 2025,
        month = jul,
          eid = {arXiv:2507.13160},
        pages = {arXiv:2507.13160},
          doi = {10.48550/arXiv.2507.13160},
archivePrefix = {arXiv},
       eprint = {2507.13160},
 primaryClass = {astro-ph.GA},
       adsurl = {https://ui.adsabs.harvard.edu/abs/2025arXiv250713160C}
}

@ARTICLE{Donnan2025,
       author = {{Donnan}, Callum T. and {Dickinson}, Mark and {Taylor}, Anthony J. and {Arrabal Haro}, Pablo and {Finkelstein}, Steven L. and {Stanton}, Thomas M. and {Jung}, Intae and {Papovich}, Casey and {Akins}, Hollis B. and {Koekemoer}, Anton M. and {McLeod}, Derek J. and {Napolitano}, Lorenzo and {Amor{\'\i}n}, Ricardo O. and {Begley}, Ryan and {Burgarella}, Denis and {Carnall}, Adam C. and {Casey}, Caitlin M. and {Calabr{\`o}}, Antonello and {Cullen}, Fergus and {Dunlop}, James S. and {Ellis}, Richard S. and {Fern{\'a}ndez}, Vital and {Giavalisco}, Mauro and {Hirschmann}, Michaela and {Hu}, Weida and {Illingworth}, Garth and {Kartaltepe}, Jeyhan S. and {Kocevski}, Dale D. and {Kokorev}, Vasily and {Leung}, Ho-Hin and {Lucas}, Ray A. and {Morales}, Alexa M. and {McLure}, Ross and {Pentericci}, Laura and {P{\'e}rez-Gonz{\'a}lez}, Pablo G. and {Somerville}, Rachel S. and {Stevenson}, Struan and {Trump}, Jonathan R. and {Yung}, L.~Y. Aaron and {Zavala}, Jorge A.},
        title = "{Very bright, very blue, and very red: JWST CAPERS analysis of highly luminous galaxies with extreme UV slopes at $\mathbf{z = 10}$}",
      journal = {arXiv e-prints},
         year = 2025,
        month = jul,
          eid = {arXiv:2507.10518},
        pages = {arXiv:2507.10518},
          doi = {10.48550/arXiv.2507.10518},
archivePrefix = {arXiv},
       eprint = {2507.10518},
 primaryClass = {astro-ph.GA},
       adsurl = {https://ui.adsabs.harvard.edu/abs/2025arXiv250710518D}
}

@ARTICLE{Tang2025,
       author = {{Tang}, Mengtao and {Stark}, Daniel P. and {Mason}, Charlotte A. and {Gelli}, Viola and {Chen}, Zuyi and {Topping}, Michael W.},
        title = "{The JWST Spectroscopic Properties of Galaxies at $z=9-14$}",
      journal = {arXiv e-prints},
         year = 2025,
        month = jul,
          eid = {arXiv:2507.08245},
        pages = {arXiv:2507.08245},
archivePrefix = {arXiv},
       eprint = {2507.08245},
 primaryClass = {astro-ph.GA},
       adsurl = {https://ui.adsabs.harvard.edu/abs/2025arXiv250708245T}
}

@ARTICLE{Pallottini2023,
       author = {{Pallottini}, A. and {Ferrara}, A.},
        title = "{Stochastic star formation in early galaxies: Implications for the James Webb Space Telescope}",
      journal = {\aap},
         year = 2023,
        month = sep,
       volume = {677},
          eid = {L4},
        pages = {L4},
          doi = {10.1051/0004-6361/202347384},
archivePrefix = {arXiv},
       eprint = {2307.03219},
 primaryClass = {astro-ph.GA},
       adsurl = {https://ui.adsabs.harvard.edu/abs/2023A&A...677L...4P}
}

@ARTICLE{Ono2023,
       author = {{Ono}, Yoshiaki and {Harikane}, Yuichi and {Ouchi}, Masami and {Yajima}, Hidenobu and {Abe}, Makito and {Isobe}, Yuki and {Shibuya}, Takatoshi and {Wise}, John H. and {Zhang}, Yechi and {Nakajima}, Kimihiko and {Umeda}, Hiroya},
        title = "{Morphologies of Galaxies at z {\ensuremath{\gtrsim}} 9 Uncovered by JWST/NIRCam Imaging: Cosmic Size Evolution and an Identification of an Extremely Compact Bright Galaxy at z 12}",
      journal = {\apj},
         year = 2023,
        month = jul,
       volume = {951},
       number = {1},
          eid = {72},
        pages = {72},
          doi = {10.3847/1538-4357/acd44a},
archivePrefix = {arXiv},
       eprint = {2208.13582},
 primaryClass = {astro-ph.GA},
       adsurl = {https://ui.adsabs.harvard.edu/abs/2023ApJ...951...72O}
}

@ARTICLE{morishita2024_sizes,
       author = {{Morishita}, Takahiro and {Stiavelli}, Massimo and {Chary}, Ranga-Ram and {Trenti}, Michele and {Bergamini}, Pietro and {Chiaberge}, Marco and {Leethochawalit}, Nicha and {Roberts-Borsani}, Guido and {Shen}, Xuejian and {Treu}, Tommaso},
        title = "{Enhanced Subkiloparsec-scale Star Formation: Results from a JWST Size Analysis of 341 Galaxies at 5 < z < 14}",
      journal = {\apj},
         year = 2024,
        month = mar,
       volume = {963},
       number = {1},
          eid = {9},
        pages = {9},
          doi = {10.3847/1538-4357/ad1404},
archivePrefix = {arXiv},
       eprint = {2308.05018},
 primaryClass = {astro-ph.GA},
       adsurl = {https://ui.adsabs.harvard.edu/abs/2024ApJ...963....9M}
}

@ARTICLE{Meena2025,
       author = {{Meena}, Ashish Kumar and {Li}, Sung Kei and {Zitrin}, Adi and {Kelly}, Patrick L. and {Broadhurst}, Tom and {Chen}, Wenlei and {Diego}, Jose M. and {Filippenko}, Alexei V. and {Furtak}, Lukas J. and {Williams}, Liliya L.~R.},
        title = "{Flashlights: Prospects for constraining the initial mass function around cosmic noon with caustic-crossing events}",
      journal = {\aap},
         year = 2025,
        month = jul,
       volume = {699},
          eid = {A299},
        pages = {A299},
          doi = {10.1051/0004-6361/202555023},
archivePrefix = {arXiv},
       eprint = {2503.21706},
 primaryClass = {astro-ph.GA},
       adsurl = {https://ui.adsabs.harvard.edu/abs/2025A&A...699A.299M}
}

@ARTICLE{jullo2007,
       author = {{Jullo}, E. and {Kneib}, J. -P. and {Limousin}, M. and {El{\'\i}asd{\'o}ttir}, {\'A}. and {Marshall}, P.~J. and {Verdugo}, T.},
        title = "{A Bayesian approach to strong lensing modelling of galaxy clusters}",
      journal = {New Journal of Physics},
         year = 2007,
        month = dec,
       volume = {9},
       number = {12},
        pages = {447},
          doi = {10.1088/1367-2630/9/12/447},
archivePrefix = {arXiv},
       eprint = {0706.0048},
 primaryClass = {astro-ph},
       adsurl = {https://ui.adsabs.harvard.edu/abs/2007NJPh....9..447J}
}

@ARTICLE{oguri2021,
       author = {{Oguri}, Masamune},
        title = "{Fast Calculation of Gravitational Lensing Properties of Elliptical Navarro-Frenk-White and Hernquist Density Profiles}",
      journal = {\pasp},
         year = 2021,
        month = jul,
       volume = {133},
       number = {1025},
          eid = {074504},
        pages = {074504},
          doi = {10.1088/1538-3873/ac12db},
archivePrefix = {arXiv},
       eprint = {2106.11464},
 primaryClass = {astro-ph.IM},
       adsurl = {https://ui.adsabs.harvard.edu/abs/2021PASP..133g4504O}
}

@ARTICLE{oguri2010,
       author = {{Oguri}, Masamune},
        title = "{The Mass Distribution of SDSS J1004+4112 Revisited}",
      journal = {\pasj},
         year = 2010,
        month = aug,
       volume = {62},
        pages = {1017},
          doi = {10.1093/pasj/62.4.1017},
archivePrefix = {arXiv},
       eprint = {1005.3103},
 primaryClass = {astro-ph.CO},
       adsurl = {https://ui.adsabs.harvard.edu/abs/2010PASJ...62.1017O}
}

@ARTICLE{Mowla2024,
       author = {{Mowla}, Lamiya and {Iyer}, Kartheik and {Asada}, Yoshihisa and {Desprez}, Guillaume and {Tan}, Vivian Yun Yan and {Martis}, Nicholas and {Sarrouh}, Ghassan and {Strait}, Victoria and {Abraham}, Roberto and {Brada{\v{c}}}, Maru{\v{s}}a and {Brammer}, Gabriel and {Muzzin}, Adam and {Pacifici}, Camilla and {Ravindranath}, Swara and {Sawicki}, Marcin and {Willott}, Chris and {Estrada-Carpenter}, Vince and {Jahan}, Nusrath and {Noirot}, Ga{\"e}l and {Matharu}, Jasleen and {Rihtar{\v{s}}i{\v{c}}}, Gregor and {Zabl}, Johannes},
        title = "{Formation of a low-mass galaxy from star clusters in a 600-million-year-old Universe}",
      journal = {\nat},
         year = 2024,
        month = dec,
       volume = {636},
       number = {8042},
        pages = {332-336},
          doi = {10.1038/s41586-024-08293-0},
archivePrefix = {arXiv},
       eprint = {2402.08696},
 primaryClass = {astro-ph.GA},
       adsurl = {https://ui.adsabs.harvard.edu/abs/2024Natur.636..332M}
}

@ARTICLE{Bradac2024,
       author = {{Brada{\v{c}}}, Maru{\v{s}}a and {Strait}, Victoria and {Mowla}, Lamiya and {Iyer}, Kartheik G. and {Noirot}, Ga{\"e}l and {Willott}, Chris and {Brammer}, Gabe and {Abraham}, Roberto and {Asada}, Yoshihisa and {Desprez}, Guillaume and {Estrada-Carpenter}, Vince and {Harshan}, Anishya and {Martis}, Nicholas S. and {Matharu}, Jasleen and {Muzzin}, Adam and {Rihtar{\v{s}}i{\v{c}}}, Gregor and {Sarrouh}, Ghassan T.~E. and {Sawicki}, Marcin},
        title = "{Star Formation at the Epoch of Reionization with CANUCS: The Ages of Stellar Populations in MACS1149-JD1}",
      journal = {\apjl},
         year = 2024,
        month = jan,
       volume = {961},
       number = {1},
          eid = {L21},
        pages = {L21},
          doi = {10.3847/2041-8213/ad0e73},
archivePrefix = {arXiv},
       eprint = {2308.13288},
 primaryClass = {astro-ph.GA},
       adsurl = {https://ui.adsabs.harvard.edu/abs/2024ApJ...961L..21B}
}

@ARTICLE{Garcia2025,
       author = {{Garcia}, Fred Angelo Batan and {Ricotti}, Massimo and {Sugimura}, Kazuyuki},
        title = "{Seeding Cores: A Pathway for Nuclear Star Clusters from Bound Star Clusters in the First Billion Years}",
      journal = {The Open Journal of Astrophysics},
         year = 2025,
        month = oct,
       volume = {8},
          eid = {146},
        pages = {146},
          doi = {10.33232/001c.145064},
archivePrefix = {arXiv},
       eprint = {2503.08779},
 primaryClass = {astro-ph.GA},
       adsurl = {https://ui.adsabs.harvard.edu/abs/2025OJAp....8E.146G}
}

@ARTICLE{Sugimura2024,
       author = {{Sugimura}, Kazuyuki and {Ricotti}, Massimo and {Park}, Jongwon and {Garcia}, Fred Angelo Batan and {Yajima}, Hidenobu},
        title = "{Violent Starbursts and Quiescence Induced by Far-ultraviolet Radiation Feedback in Metal-poor Galaxies at High Redshift}",
      journal = {\apj},
         year = 2024,
        month = jul,
       volume = {970},
       number = {1},
          eid = {14},
        pages = {14},
          doi = {10.3847/1538-4357/ad499a},
archivePrefix = {arXiv},
       eprint = {2403.04824},
 primaryClass = {astro-ph.GA},
       adsurl = {https://ui.adsabs.harvard.edu/abs/2024ApJ...970...14S}
}

@ARTICLE{Garcia2023,
       author = {{Garcia}, Fred Angelo Batan and {Ricotti}, Massimo and {Sugimura}, Kazuyuki and {Park}, Jongwon},
        title = "{Star cluster formation and survival in the first galaxies}",
      journal = {\mnras},
         year = 2023,
        month = jun,
       volume = {522},
       number = {2},
        pages = {2495-2515},
          doi = {10.1093/mnras/stad1092},
archivePrefix = {arXiv},
       eprint = {2212.13946},
 primaryClass = {astro-ph.GA},
       adsurl = {https://ui.adsabs.harvard.edu/abs/2023MNRAS.522.2495G}
}

@ARTICLE{Steinhardt23,
       author = {{Steinhardt}, Charles L. and {Kokorev}, Vasily and {Rusakov}, Vadim and {Garcia}, Ethan and {Sneppen}, Albert},
        title = "{Templates for Fitting Photometry of Ultra-high-redshift Galaxies}",
      journal = {\apjl},
         year = 2023,
        month = jul,
       volume = {951},
       number = {2},
          eid = {L40},
        pages = {L40},
          doi = {10.3847/2041-8213/acdef6},
archivePrefix = {arXiv},
       eprint = {2208.07879},
 primaryClass = {astro-ph.GA},
       adsurl = {https://ui.adsabs.harvard.edu/abs/2023ApJ...951L..40S}
}

@ARTICLE{Chon21,
       author = {{Chon}, Sunmyon and {Omukai}, Kazuyuki and {Schneider}, Raffaella},
        title = "{Transition of the initial mass function in the metal-poor environments}",
      journal = {\mnras},
         year = 2021,
        month = dec,
       volume = {508},
       number = {3},
        pages = {4175-4192},
          doi = {10.1093/mnras/stab2497},
archivePrefix = {arXiv},
       eprint = {2103.04997},
 primaryClass = {astro-ph.GA},
       adsurl = {https://ui.adsabs.harvard.edu/abs/2021MNRAS.508.4175C}
}

@ARTICLE{Messa25_CG,
       author = {{Messa}, M. and {Vanzella}, E. and {Loiacono}, F. and {Adamo}, A. and {Oguri}, M. and {Sharon}, K. and {Bradley}, L.~D. and {Christensen}, L. and {Claeyssens}, A. and {Richard}, J. and {Abdurro'uf} and {Bauer}, F.~E. and {Bergamini}, P. and {Bolamperti}, A. and {Brada{\v{c}}}, M. and {Calura}, F. and {Coe}, D. and {Diego}, J.~M. and {Grillo}, C. and {Y-Y. Hsiao}, T. and {Inoue}, A.~K. and {Fujimoto}, S. and {Lombardi}, M. and {Meneghetti}, M. and {Resseguier}, T. and {Ricotti}, M. and {Rosati}, P. and {Welch}, B. and {Windhorst}, R.~A. and {Xu}, X. and {Zackrisson}, E. and {Zanella}, A. and {Zitrin}, A.},
        title = "{JWST Spectroscopic Confirmation of the Cosmic Gems Arc at z=9.625 -- Insights into the small scale structure of a post-burst system}",
      journal = {arXiv e-prints},
         year = 2025,
        month = jul,
          eid = {arXiv:2507.18705},
        pages = {arXiv:2507.18705},
          doi = {10.48550/arXiv.2507.18705},
archivePrefix = {arXiv},
       eprint = {2507.18705},
 primaryClass = {astro-ph.GA},
       adsurl = {https://ui.adsabs.harvard.edu/abs/2025arXiv250718705M}
}

@ARTICLE{Ji2025,
       author = {{Ji}, Xihan and {Belokurov}, Vasily and {Maiolino}, Roberto and {Monty}, Stephanie and {Isobe}, Yuki and {Kravtsov}, Andrey and {McClymont}, William and {{\"U}bler}, Hannah},
        title = "{Connecting JWST discovered N/O-enhanced galaxies to globular clusters: Evidence from chemical imprints}",
      journal = {arXiv e-prints},
         year = 2025,
        month = may,
          eid = {arXiv:2505.12505},
        pages = {arXiv:2505.12505},
archivePrefix = {arXiv},
       eprint = {2505.12505},
 primaryClass = {astro-ph.GA},
       adsurl = {https://ui.adsabs.harvard.edu/abs/2025arXiv250512505J}
}

@ARTICLE{Naidu2025,
       author = {{Naidu}, Rohan P. and {Oesch}, Pascal A. and {Brammer}, Gabriel and {Weibel}, Andrea and {Li}, Yijia and {Matthee}, Jorryt and {Chisholm}, John and {Pollock}, Clara L. and {Heintz}, Kasper E. and {Johnson}, Benjamin D. and {Shen}, Xuejian and {Hviding}, Raphael E. and {Leja}, Joel and {Tacchella}, Sandro and {Ganguly}, Arpita and {Witten}, Callum and {Atek}, Hakim and {Belli}, Sirio and {Bose}, Sownak and {Bouwens}, Rychard and {Dayal}, Pratika and {Decarli}, Roberto and {de Graaff}, Anna and {Fudamoto}, Yoshinobu and {Giovinazzo}, Emma and {Greene}, Jenny E. and {Illingworth}, Garth and {Inoue}, Akio K. and {Kane}, Sarah G. and {Labbe}, Ivo and {Leonova}, Ecaterina and {Marques-Chaves}, Rui and {Meyer}, Romain A. and {Nelson}, Erica J. and {Roberts-Borsani}, Guido and {Schaerer}, Daniel and {Simcoe}, Robert A. and {Stefanon}, Mauro and {Sugahara}, Yuma and {Toft}, Sune and {van der Wel}, Arjen and {van Dokkum}, Pieter and {Walter}, Fabian and {Watson}, Darach and {Weaver}, John R. and {Whitaker}, Katherine E.},
        title = "{A Cosmic Miracle: A Remarkably Luminous Galaxy at $z_{\rm{spec}}=14.44$ Confirmed with JWST}",
      journal = {arXiv e-prints},
         year = 2025,
        month = may,
          eid = {arXiv:2505.11263},
        pages = {arXiv:2505.11263},
archivePrefix = {arXiv},
       eprint = {2505.11263},
 primaryClass = {astro-ph.GA},
       adsurl = {https://ui.adsabs.harvard.edu/abs/2025arXiv250511263N}
}

@ARTICLE{Dekel23,
       author = {{Dekel}, Avishai and {Sarkar}, Kartick C. and {Birnboim}, Yuval and {Mandelker}, Nir and {Li}, Zhaozhou},
        title = "{Efficient formation of massive galaxies at cosmic dawn by feedback-free starbursts}",
      journal = {\mnras},
         year = 2023,
        month = aug,
       volume = {523},
       number = {3},
        pages = {3201-3218},
          doi = {10.1093/mnras/stad1557},
archivePrefix = {arXiv},
       eprint = {2303.04827},
 primaryClass = {astro-ph.GA},
       adsurl = {https://ui.adsabs.harvard.edu/abs/2023MNRAS.523.3201D}
}

@ARTICLE{Pascale25,
       author = {{Pascale}, R. and {Calura}, F. and {Vesperini}, E. and {Rosdahl}, J. and {Nipoti}, C. and {Giunchi}, E. and {Lacchin}, E. and {Lupi}, A. and {Messa}, M. and {Meneghetti}, M. and {Ragagnin}, A. and {Vanzella}, E. and {Zanella}, A.},
        title = "{SIEGE: IV. Compact star clusters in cosmological simulations with a high star formation efficiency and subparsec resolution}",
      journal = {\aap},
         year = 2025,
        month = jun,
       volume = {699},
          eid = {A31},
        pages = {A31},
          doi = {10.1051/0004-6361/202453252},
archivePrefix = {arXiv},
       eprint = {2505.06346},
 primaryClass = {astro-ph.GA},
       adsurl = {https://ui.adsabs.harvard.edu/abs/2025A&A...699A..31P}
}

@ARTICLE{linden2021,
       author = {{Linden}, S.~T. and {Evans}, A.~S. and {Larson}, K. and {Privon}, G.~C. and {Armus}, L. and {Rich}, J. and {D{\'\i}az-Santos}, T. and {Murphy}, E.~J. and {Song}, Y. and {Barcos-Mu{\~n}oz}, L. and {Howell}, J. and {Charmandaris}, V. and {Inami}, H. and {U}, V. and {Surace}, J.~A. and {Mazzarella}, J.~M. and {Calzetti}, D.},
        title = "{Massive Star Cluster Formation and Destruction in Luminous Infrared Galaxies in GOALS. II. An ACS/WFC3 Survey of Nearby LIRGs}",
      journal = {\apj},
         year = 2021,
        month = dec,
       volume = {923},
       number = {2},
          eid = {278},
        pages = {278},
          doi = {10.3847/1538-4357/ac2892},
archivePrefix = {arXiv},
       eprint = {2110.03638},
 primaryClass = {astro-ph.GA},
       adsurl = {https://ui.adsabs.harvard.edu/abs/2021ApJ...923..278L}
}

@ARTICLE{Eldridge2017,
       author = {{Eldridge}, J.~J. and {Stanway}, E.~R. and {Xiao}, L. and {McClelland}, L.~A.~S. and {Taylor}, G. and {Ng}, M. and {Greis}, S.~M.~L. and {Bray}, J.~C.},
        title = "{Binary Population and Spectral Synthesis Version 2.1: Construction, Observational Verification, and New Results}",
      journal = {\pasa},
         year = 2017,
        month = nov,
       volume = {34},
          eid = {e058},
        pages = {e058},
          doi = {10.1017/pasa.2017.51},
archivePrefix = {arXiv},
       eprint = {1710.02154},
 primaryClass = {astro-ph.SR},
       adsurl = {https://ui.adsabs.harvard.edu/abs/2017PASA...34...58E}
}

@ARTICLE{Plank18,
       author = {{Planck Collaboration} and {Akrami}, Y. and {Arg{\"u}eso}, F. and {Ashdown}, M. and {Aumont}, J. and {Baccigalupi}, C. and {Ballardini}, M. and {Banday}, A.~J. and {Barreiro}, R.~B. and {Bartolo}, N. and {Basak}, S. and {Benabed}, K. and {Bernard}, J. -P. and {Bersanelli}, M. and {Bielewicz}, P. and {Bonavera}, L. and {Bond}, J.~R. and {Borrill}, J. and {Bouchet}, F.~R. and {Burigana}, C. and {Butler}, R.~C. and {Calabrese}, E. and {Carron}, J. and {Chiang}, H.~C. and {Combet}, C. and {Crill}, B.~P. and {Cuttaia}, F. and {de Bernardis}, P. and {de Rosa}, A. and {de Zotti}, G. and {Delabrouille}, J. and {Delouis}, J. -M. and {Di Valentino}, E. and {Dickinson}, C. and {Diego}, J.~M. and {Ducout}, A. and {Dupac}, X. and {Efstathiou}, G. and {Elsner}, F. and {En{\ss}lin}, T.~A. and {Eriksen}, H.~K. and {Fantaye}, Y. and {Finelli}, F. and {Frailis}, M. and {Fraisse}, A.~A. and {Franceschi}, E. and {Frolov}, A. and {Galeotta}, S. and {Galli}, S. and {Ganga}, K. and {G{\'e}nova-Santos}, R.~T. and {Gerbino}, M. and {Ghosh}, T. and {Gonz{\'a}lez-Nuevo}, J. and {G{\'o}rski}, K.~M. and {Gratton}, S. and {Gruppuso}, A. and {Gudmundsson}, J.~E. and {Handley}, W. and {Hansen}, F.~K. and {Herranz}, D. and {Hivon}, E. and {Huang}, Z. and {Jaffe}, A.~H. and {Jones}, W.~C. and {Keih{\"a}nen}, E. and {Keskitalo}, R. and {Kiiveri}, K. and {Kim}, J. and {Kisner}, T.~S. and {Krachmalnicoff}, N. and {Kunz}, M. and {Kurki-Suonio}, H. and {L{\"a}hteenm{\"a}ki}, A. and {Lamarre}, J. -M. and {Lasenby}, A. and {Lattanzi}, M. and {Lawrence}, C.~R. and {Levrier}, F. and {Liguori}, M. and {Lilje}, P.~B. and {Lindholm}, V. and {L{\'o}pez-Caniego}, M. and {Ma}, Y. -Z. and {Mac{\'\i}as-P{\'e}rez}, J.~F. and {Maggio}, G. and {Maino}, D. and {Mandolesi}, N. and {Mangilli}, A. and {Maris}, M. and {Martin}, P.~G. and {Mart{\'\i}nez-Gonz{\'a}lez}, E. and {Matarrese}, S. and {McEwen}, J.~D. and {Meinhold}, P.~R. and {Melchiorri}, A. and {Mennella}, A. and {Migliaccio}, M. and {Miville-Desch{\^e}nes}, M. -A. and {Molinari}, D. and {Moneti}, A. and {Montier}, L. and {Morgante}, G. and {Natoli}, P. and {Oxborrow}, C.~A. and {Pagano}, L. and {Paoletti}, D. and {Partridge}, B. and {Patanchon}, G. and {Pearson}, T.~J. and {Pettorino}, V. and {Piacentini}, F. and {Polenta}, G. and {Puget}, J. -L. and {Rachen}, J.~P. and {Racine}, B. and {Reinecke}, M. and {Remazeilles}, M. and {Renzi}, A. and {Rocha}, G. and {Roudier}, G. and {Rubi{\~n}o-Mart{\'\i}n}, J.~A. and {Salvati}, L. and {Sandri}, M. and {Savelainen}, M. and {Scott}, D. and {Suur-Uski}, A. -S. and {Tauber}, J.~A. and {Tavagnacco}, D. and {Toffolatti}, L. and {Tomasi}, M. and {Trombetti}, T. and {Tucci}, M. and {Valiviita}, J. and {Van Tent}, B. and {Vielva}, P. and {Villa}, F. and {Vittorio}, N. and {Wehus}, I.~K. and {Zacchei}, A. and {Zonca}, A.},
        title = "{Planck intermediate results. LIV. The Planck multi-frequency catalogue of non-thermal sources}",
      journal = {\aap},
         year = 2018,
        month = nov,
       volume = {619},
          eid = {A94},
        pages = {A94},
          doi = {10.1051/0004-6361/201832888},
archivePrefix = {arXiv},
       eprint = {1802.08649},
 primaryClass = {astro-ph.CO},
       adsurl = {https://ui.adsabs.harvard.edu/abs/2018A&A...619A..94P}
}

@INPROCEEDINGS{Kruijssen2025,
       author = {{Kruijssen}, J.~M. Diederik},
        title = "{The formation of globular clusters}",
    booktitle = {Encyclopedia of Astrophysics, Volume 4},
         year = 2026,
       volume = {4},
        month = jan,
        pages = {500-534},
          doi = {10.1016/B978-0-443-21439-4.00078-X},
archivePrefix = {arXiv},
       eprint = {2501.16438},
 primaryClass = {astro-ph.GA},
       adsurl = {https://ui.adsabs.harvard.edu/abs/2026enap....4..500K}
}

@ARTICLE{Rivera2024,
       author = {{Rivera-Thorsen}, T. Emil and {Chisholm}, J. and {Welch}, B. and {Rigby}, J.~R. and {Hutchison}, T. and {Florian}, M. and {Sharon}, K. and {Choe}, S. and {Dahle}, H. and {Bayliss}, M.~B. and {Khullar}, G. and {Gladders}, M. and {Hayes}, M. and {Adamo}, A. and {Owens}, M.~R. and {Kim}, K.},
        title = "{The Sunburst Arc with JWST: I. Detection of Wolf-Rayet stars injecting nitrogen into a low-metallicity, z = 2.37 proto-globular cluster leaking ionizing photons}",
      journal = {\aap},
         year = 2024,
        month = oct,
       volume = {690},
          eid = {A269},
        pages = {A269},
          doi = {10.1051/0004-6361/202450359},
archivePrefix = {arXiv},
       eprint = {2404.08884},
 primaryClass = {astro-ph.GA},
       adsurl = {https://ui.adsabs.harvard.edu/abs/2024A&A...690A.269R}
}

@ARTICLE{Elmegreen2006_IMFstellar_cluster,
       author = {{Elmegreen}, Bruce G.},
        title = "{On the Similarity between Cluster and Galactic Stellar Initial Mass Functions}",
      journal = {\apj},
         year = 2006,
        month = sep,
       volume = {648},
       number = {1},
        pages = {572-579},
          doi = {10.1086/505785},
archivePrefix = {arXiv},
       eprint = {astro-ph/0605520},
 primaryClass = {astro-ph},
       adsurl = {https://ui.adsabs.harvard.edu/abs/2006ApJ...648..572E}
}

@ARTICLE{Andersson2024_preSN,
       author = {{Andersson}, Eric P. and {Mac Low}, Mordecai-Mark and {Agertz}, Oscar and {Renaud}, Florent and {Li}, Hui},
        title = "{Pre-supernova feedback sets the star cluster mass function to a power law and reduces the cluster formation efficiency}",
      journal = {\aap},
         year = 2024,
        month = jan,
       volume = {681},
          eid = {A28},
        pages = {A28},
          doi = {10.1051/0004-6361/202347792},
archivePrefix = {arXiv},
       eprint = {2308.12363},
 primaryClass = {astro-ph.GA},
       adsurl = {https://ui.adsabs.harvard.edu/abs/2024A&A...681A..28A}
}

@ARTICLE{Nebrin2025,
       author = {{Nebrin}, Olof and {Smith}, Aaron and {Lorinc}, Kevin and {H{\"o}rnquist}, Johan and {Larson}, {\r{A}}sa and {Mellema}, Garrelt and {Giri}, Sambit K.},
        title = "{Lyman-{\ensuremath{\alpha}} feedback prevails at Cosmic Dawn: implications for the first galaxies, stars, and star clusters}",
      journal = {\mnras},
         year = 2025,
        month = feb,
       volume = {537},
       number = {2},
        pages = {1646-1687},
          doi = {10.1093/mnras/staf038},
archivePrefix = {arXiv},
       eprint = {2409.19288},
 primaryClass = {astro-ph.GA},
       adsurl = {https://ui.adsabs.harvard.edu/abs/2025MNRAS.537.1646N}
}

@ARTICLE{Renzini2025,
       author = {{Renzini}, Alvio},
        title = "{On the ubiquity of extreme baryon concentrations in the early Universe}",
      journal = {\mnras},
         year = 2025,
        month = jan,
       volume = {536},
       number = {1},
        pages = {L8-L12},
          doi = {10.1093/mnrasl/slae101},
archivePrefix = {arXiv},
       eprint = {2410.22138},
 primaryClass = {astro-ph.GA},
       adsurl = {https://ui.adsabs.harvard.edu/abs/2025MNRAS.536L...8R}
}

@ARTICLE{cook2023,
       author = {{Cook}, D.~O. and {Lee}, J.~C. and {Adamo}, A. and {Calzetti}, D. and {Chandar}, R. and {Whitmore}, B.~C. and {Aloisi}, A. and {Cignoni}, M. and {Dale}, D.~A. and {Elmegreen}, B.~G. and {Fumagalli}, M. and {Grasha}, K. and {Johnson}, K.~E. and {Kennicutt}, R.~C. and {Kim}, H. and {Linden}, S.~T. and {Messa}, M. and {{\"O}stlin}, G. and {Ryon}, J.~E. and {Sacchi}, E. and {Thilker}, D.~A. and {Tosi}, M. and {Wofford}, A.},
        title = "{Fraction of stars in clusters for the LEGUS dwarf galaxies}",
      journal = {\mnras},
         year = 2023,
        month = mar,
       volume = {519},
       number = {3},
        pages = {3749-3775},
          doi = {10.1093/mnras/stac3748},
archivePrefix = {arXiv},
       eprint = {2212.07519},
 primaryClass = {astro-ph.GA},
       adsurl = {https://ui.adsabs.harvard.edu/abs/2023MNRAS.519.3749C}
}

@ARTICLE{Nakazato2024,
       author = {{Nakazato}, Yurina and {Ferrara}, Andrea},
        title = "{Radiation-driven dusty outflows from early galaxies}",
      journal = {arXiv e-prints},
         year = 2024,
        month = dec,
          eid = {arXiv:2412.07598},
        pages = {arXiv:2412.07598},
          doi = {10.48550/arXiv.2412.07598},
archivePrefix = {arXiv},
       eprint = {2412.07598},
 primaryClass = {astro-ph.GA},
       adsurl = {https://ui.adsabs.harvard.edu/abs/2024arXiv241207598N}
}

@ARTICLE{Whitler2025,
       author = {{Whitler}, Lily and {Stark}, Daniel P. and {Topping}, Michael W. and {Robertson}, Brant and {Rieke}, Marcia and {Hainline}, Kevin N. and {Endsley}, Ryan and {Chen}, Zuyi and {Baker}, William M. and {Bhatawdekar}, Rachana and {Bunker}, Andrew J. and {Carniani}, Stefano and {Charlot}, St{\'e}phane and {Chevallard}, Jacopo and {Curtis-Lake}, Emma and {Egami}, Eiichi and {Eisenstein}, Daniel J. and {Helton}, Jakob M. and {Ji}, Zhiyuan and {Johnson}, Benjamin D. and {P{\'e}rez-Gonz{\'a}lez}, Pablo G. and {Rinaldi}, Pierluigi and {Tacchella}, Sandro and {Williams}, Christina C. and {Willmer}, Christopher N.~A. and {Willott}, Chris and {Witstok}, Joris},
        title = "{The z {\ensuremath{\gtrsim}} 9 Galaxy UV Luminosity Function from the JWST Advanced Deep Extragalactic Survey: Insights into Early Galaxy Evolution and Reionization}",
      journal = {\apj},
         year = 2025,
        month = oct,
       volume = {992},
       number = {1},
          eid = {63},
        pages = {63},
          doi = {10.3847/1538-4357/adfddc},
archivePrefix = {arXiv},
       eprint = {2501.00984},
 primaryClass = {astro-ph.GA},
       adsurl = {https://ui.adsabs.harvard.edu/abs/2025ApJ...992...63W}
}

@ARTICLE{Calura2025,
       author = {{Calura}, F. and {Pascale}, R. and {Agertz}, O. and {Andersson}, E. and {Lacchin}, E. and {Lupi}, A. and {Meneghetti}, M. and {Nipoti}, C. and {Ragagnin}, A. and {Rosdahl}, J. and {Vanzella}, E. and {Vesperini}, E. and {Zanella}, A.},
        title = "{SIEGE: III. The formation of dense stellar clusters in sub-parsec resolution cosmological simulations with individual star feedback}",
      journal = {\aap},
         year = 2025,
        month = jun,
       volume = {698},
          eid = {A207},
        pages = {A207},
          doi = {10.1051/0004-6361/202452876},
archivePrefix = {arXiv},
       eprint = {2411.02502},
 primaryClass = {astro-ph.GA},
       adsurl = {https://ui.adsabs.harvard.edu/abs/2025A&A...698A.207C}
}

@ARTICLE{Trujillo-Gomez19,
       author = {{Trujillo-Gomez}, Sebastian and {Reina-Campos}, Marta and {Kruijssen}, J.~M. Diederik},
        title = "{A model for the minimum mass of bound stellar clusters and its dependence on the galactic environment}",
      journal = {\mnras},
         year = 2019,
        month = sep,
       volume = {488},
       number = {3},
        pages = {3972-3994},
          doi = {10.1093/mnras/stz1932},
archivePrefix = {arXiv},
       eprint = {1907.04861},
 primaryClass = {astro-ph.GA},
       adsurl = {https://ui.adsabs.harvard.edu/abs/2019MNRAS.488.3972T}
}

@ARTICLE{Perez-Gonzalez2023,
       author = {{P{\'e}rez-Gonz{\'a}lez}, Pablo G. and {Costantin}, Luca and {Langeroodi}, Danial and {Rinaldi}, Pierluigi and {Annunziatella}, Marianna and {Ilbert}, Olivier and {Colina}, Luis and {N{\o}rgaard-Nielsen}, Hans Ulrik and {Greve}, Thomas R. and {{\"O}stlin}, G{\"o}ran and {Wright}, Gillian and {Alonso-Herrero}, Almudena and {{\'A}lvarez-M{\'a}rquez}, Javier and {Caputi}, Karina I. and {Eckart}, Andreas and {Le F{\`e}vre}, Olivier and {Labiano}, {\'A}lvaro and {Garc{\'\i}a-Mar{\'\i}n}, Macarena and {Hjorth}, Jens and {Kendrew}, Sarah and {Pye}, John P. and {Tikkanen}, Tuomo and {van der Werf}, Paul and {Walter}, Fabian and {Ward}, Martin and {Bik}, Arjan and {Boogaard}, Leindert and {Bosman}, Sarah E.~I. and {G{\'o}mez}, Alejandro Crespo and {Gillman}, Steven and {Iani}, Edoardo and {Jermann}, Iris and {Melinder}, Jens and {Meyer}, Romain A. and {Moutard}, Thibaud and {van Dishoek}, Ewine and {Henning}, Thomas and {Lagage}, Pierre-Olivier and {Guedel}, Manuel and {Peissker}, Florian and {Ray}, Tom and {Vandenbussche}, Bart and {Garc{\'\i}a-Argum{\'a}nez}, {\'A}ngela and {Mar{\'\i}a M{\'e}rida}, Rosa},
        title = "{Life beyond 30: Probing the -20 < M $_{UV}$ < -17 Luminosity Function at 8 < z < 13 with the NIRCam Parallel Field of the MIRI Deep Survey}",
      journal = {\apjl},
         year = 2023,
        month = jul,
       volume = {951},
       number = {1},
          eid = {L1},
        pages = {L1},
          doi = {10.3847/2041-8213/acd9d0},
archivePrefix = {arXiv},
       eprint = {2302.02429},
 primaryClass = {astro-ph.GA},
       adsurl = {https://ui.adsabs.harvard.edu/abs/2023ApJ...951L...1P}
}

@ARTICLE{McLeod2024,
       author = {{McLeod}, D.~J. and {Donnan}, C.~T. and {McLure}, R.~J. and {Dunlop}, J.~S. and {Magee}, D. and {Begley}, R. and {Carnall}, A.~C. and {Cullen}, F. and {Ellis}, R.~S. and {Hamadouche}, M.~L. and {Stanton}, T.~M.},
        title = "{The galaxy UV luminosity function at z ≃ 11 from a suite of public JWST ERS, ERO, and Cycle-1 programs}",
      journal = {\mnras},
         year = 2024,
        month = jan,
       volume = {527},
       number = {3},
        pages = {5004-5022},
          doi = {10.1093/mnras/stad3471},
archivePrefix = {arXiv},
       eprint = {2304.14469},
 primaryClass = {astro-ph.GA},
       adsurl = {https://ui.adsabs.harvard.edu/abs/2024MNRAS.527.5004M}
}

@ARTICLE{Donnan2024,
       author = {{Donnan}, C.~T. and {McLure}, R.~J. and {Dunlop}, J.~S. and {McLeod}, D.~J. and {Magee}, D. and {Arellano-C{\'o}rdova}, K.~Z. and {Barrufet}, L. and {Begley}, R. and {Bowler}, R.~A.~A. and {Carnall}, A.~C. and {Cullen}, F. and {Ellis}, R.~S. and {Fontana}, A. and {Illingworth}, G.~D. and {Grogin}, N.~A. and {Hamadouche}, M.~L. and {Koekemoer}, A.~M. and {Liu}, F. -Y. and {Mason}, C. and {Santini}, P. and {Stanton}, T.~M.},
        title = "{JWST PRIMER: a new multifield determination of the evolving galaxy UV luminosity function at redshifts z ≃ 9 - 15}",
      journal = {\mnras},
         year = 2024,
        month = sep,
       volume = {533},
       number = {3},
        pages = {3222-3237},
          doi = {10.1093/mnras/stae2037},
archivePrefix = {arXiv},
       eprint = {2403.03171},
 primaryClass = {astro-ph.GA},
       adsurl = {https://ui.adsabs.harvard.edu/abs/2024MNRAS.533.3222D}
}

@ARTICLE{Castellano2023,
       author = {{Castellano}, Marco and {Fontana}, Adriano and {Treu}, Tommaso and {Merlin}, Emiliano and {Santini}, Paola and {Bergamini}, Pietro and {Grillo}, Claudio and {Rosati}, Piero and {Acebron}, Ana and {Leethochawalit}, Nicha and {Paris}, Diego and {Bonchi}, Andrea and {Belfiori}, Davide and {Calabr{\`o}}, Antonello and {Correnti}, Matteo and {Nonino}, Mario and {Polenta}, Gianluca and {Trenti}, Michele and {Boyett}, Kristan and {Brammer}, G. and {Broadhurst}, Tom and {Caminha}, Gabriel B. and {Chen}, Wenlei and {Filippenko}, Alexei V. and {Fortuni}, Flaminia and {Glazebrook}, Karl and {Mascia}, Sara and {Mason}, Charlotte A. and {Menci}, Nicola and {Meneghetti}, Massimo and {Mercurio}, Amata and {Metha}, Benjamin and {Morishita}, Takahiro and {Nanayakkara}, Themiya and {Pentericci}, Laura and {Roberts-Borsani}, Guido and {Roy}, Namrata and {Vanzella}, Eros and {Vulcani}, Benedetta and {Yang}, Lilan and {Wang}, Xin},
        title = "{Early Results from GLASS-JWST. XIX. A High Density of Bright Galaxies at z {\ensuremath{\approx}} 10 in the A2744 Region}",
      journal = {\apjl},
         year = 2023,
        month = may,
       volume = {948},
       number = {2},
          eid = {L14},
        pages = {L14},
          doi = {10.3847/2041-8213/accea5},
archivePrefix = {arXiv},
       eprint = {2212.06666},
 primaryClass = {astro-ph.GA},
       adsurl = {https://ui.adsabs.harvard.edu/abs/2023ApJ...948L..14C}
}

@ARTICLE{Finkelstein2024,
       author = {{Finkelstein}, Steven L. and {Leung}, Gene C.~K. and {Bagley}, Micaela B. and {Dickinson}, Mark and {Ferguson}, Henry C. and {Papovich}, Casey and {Akins}, Hollis B. and {Arrabal Haro}, Pablo and {Dav{\'e}}, Romeel and {Dekel}, Avishai and {Kartaltepe}, Jeyhan S. and {Kocevski}, Dale D. and {Koekemoer}, Anton M. and {Pirzkal}, Nor and {Somerville}, Rachel S. and {Yung}, L.~Y. Aaron and {Amor{\'\i}n}, Ricardo O. and {Backhaus}, Bren E. and {Behroozi}, Peter and {Bisigello}, Laura and {Bromm}, Volker and {Casey}, Caitlin M. and {Ch{\'a}vez Ortiz}, {\'O}scar A. and {Cheng}, Yingjie and {Chworowsky}, Katherine and {Cleri}, Nikko J. and {Cooper}, M.~C. and {Davis}, Kelcey and {de la Vega}, Alexander and {Elbaz}, David and {Franco}, Maximilien and {Fontana}, Adriano and {Fujimoto}, Seiji and {Giavalisco}, Mauro and {Grogin}, Norman A. and {Holwerda}, Benne W. and {Huertas-Company}, Marc and {Hirschmann}, Michaela and {Iyer}, Kartheik G. and {Jogee}, Shardha and {Jung}, Intae and {Larson}, Rebecca L. and {Lucas}, Ray A. and {Mobasher}, Bahram and {Morales}, Alexa M. and {Morley}, Caroline V. and {Mukherjee}, Sagnick and {P{\'e}rez-Gonz{\'a}lez}, Pablo G. and {Ravindranath}, Swara and {Rodighiero}, Giulia and {Rowland}, Melanie J. and {Tacchella}, Sandro and {Taylor}, Anthony J. and {Trump}, Jonathan R. and {Wilkins}, Stephen M.},
        title = "{The Complete CEERS Early Universe Galaxy Sample: A Surprisingly Slow Evolution of the Space Density of Bright Galaxies at z {\ensuremath{\sim}} 8.5{\textendash}14.5}",
      journal = {\apjl},
         year = 2024,
        month = jul,
       volume = {969},
       number = {1},
          eid = {L2},
        pages = {L2},
          doi = {10.3847/2041-8213/ad4495},
archivePrefix = {arXiv},
       eprint = {2311.04279},
 primaryClass = {astro-ph.GA},
       adsurl = {https://ui.adsabs.harvard.edu/abs/2024ApJ...969L...2F}
}

@ARTICLE{Napolitano2024,
       author = {{Napolitano}, L. and {Castellano}, M. and {Pentericci}, L. and {Arrabal Haro}, P. and {Fontana}, A. and {Treu}, T. and {Bergamini}, P. and {Calabr{\`o}}, A. and {Mascia}, S. and {Morishita}, T. and {Roberts-Borsani}, G. and {Santini}, P. and {Vanzella}, E. and {Vulcani}, B. and {Zakharova}, D. and {Bakx}, T. and {Dickinson}, M. and {Grillo}, C. and {Leethochawalit}, N. and {Llerena}, M. and {Merlin}, E. and {Paris}, D. and {Rojas-Ruiz}, S. and {Rosati}, P. and {Wang}, X. and {Yoon}, I. and {Zavala}, J.},
        title = "{Seven wonders of Cosmic Dawn: JWST confirms a high abundance of galaxies and AGN at z ≃ 9{\textendash}11 in the GLASS field}",
      journal = {\aap},
         year = 2025,
        month = jan,
       volume = {693},
          eid = {A50},
        pages = {A50},
          doi = {10.1051/0004-6361/202452090},
archivePrefix = {arXiv},
       eprint = {2410.10967},
 primaryClass = {astro-ph.GA},
       adsurl = {https://ui.adsabs.harvard.edu/abs/2025A&A...693A..50N}
}

@ARTICLE{Harikane2024,
       author = {{Harikane}, Yuichi and {Nakajima}, Kimihiko and {Ouchi}, Masami and {Umeda}, Hiroya and {Isobe}, Yuki and {Ono}, Yoshiaki and {Xu}, Yi and {Zhang}, Yechi},
        title = "{Pure Spectroscopic Constraints on UV Luminosity Functions and Cosmic Star Formation History from 25 Galaxies at z $_{spec}$ = 8.61-13.20 Confirmed with JWST/NIRSpec}",
      journal = {\apj},
         year = 2024,
        month = jan,
       volume = {960},
       number = {1},
          eid = {56},
        pages = {56},
          doi = {10.3847/1538-4357/ad0b7e},
archivePrefix = {arXiv},
       eprint = {2304.06658},
 primaryClass = {astro-ph.GA},
       adsurl = {https://ui.adsabs.harvard.edu/abs/2024ApJ...960...56H}
}

@ARTICLE{Harikane2023,
       author = {{Harikane}, Yuichi and {Ouchi}, Masami and {Oguri}, Masamune and {Ono}, Yoshiaki and {Nakajima}, Kimihiko and {Isobe}, Yuki and {Umeda}, Hiroya and {Mawatari}, Ken and {Zhang}, Yechi},
        title = "{A Comprehensive Study of Galaxies at z   9-16 Found in the Early JWST Data: Ultraviolet Luminosity Functions and Cosmic Star Formation History at the Pre-reionization Epoch}",
      journal = {\apjs},
         year = 2023,
        month = mar,
       volume = {265},
       number = {1},
          eid = {5},
        pages = {5},
          doi = {10.3847/1538-4365/acaaa9},
archivePrefix = {arXiv},
       eprint = {2208.01612},
 primaryClass = {astro-ph.GA},
       adsurl = {https://ui.adsabs.harvard.edu/abs/2023ApJS..265....5H}
}

@ARTICLE{adamo2020SSRv,
       author = {{Adamo}, Angela and {Zeidler}, Peter and {Kruijssen}, J.~M. Diederik and {Chevance}, M{\'e}lanie and {Gieles}, Mark and {Calzetti}, Daniela and {Charbonnel}, Corinne and {Zinnecker}, Hans and {Krause}, Martin G.~H.},
        title = "{Star Clusters Near and Far; Tracing Star Formation Across Cosmic Time}",
      journal = {\ssr},
         year = 2020,
        month = jun,
       volume = {216},
       number = {4},
          eid = {69},
        pages = {69},
          doi = {10.1007/s11214-020-00690-x},
archivePrefix = {arXiv},
       eprint = {2005.06188},
 primaryClass = {astro-ph.GA},
       adsurl = {https://ui.adsabs.harvard.edu/abs/2020SSRv..216...69A}
}

@ARTICLE{messa2018,
       author = {{Messa}, M. and {Adamo}, A. and {{\"O}stlin}, G. and {Calzetti}, D. and {Grasha}, K. and {Grebel}, E.~K. and {Shabani}, F. and {Chandar}, R. and {Dale}, D.~A. and {Dobbs}, C.~L. and {Elmegreen}, B.~G. and {Fumagalli}, M. and {Gouliermis}, D.~A. and {Kim}, H. and {Smith}, L.~J. and {Thilker}, D.~A. and {Tosi}, M. and {Ubeda}, L. and {Walterbos}, R. and {Whitmore}, B.~C. and {Fedorenko}, K. and {Mahadevan}, S. and {Andrews}, J.~E. and {Bright}, S.~N. and {Cook}, D.~O. and {Kahre}, L. and {Nair}, P. and {Pellerin}, A. and {Ryon}, J.~E. and {Ahmad}, S.~D. and {Beale}, L.~P. and {Brown}, K. and {Clarkson}, D.~A. and {Guidarelli}, G.~C. and {Parziale}, R. and {Turner}, J. and {Weber}, M.},
        title = "{The young star cluster population of M51 with LEGUS - I. A comprehensive study of cluster formation and evolution}",
      journal = {\mnras},
         year = 2018,
        month = jan,
       volume = {473},
       number = {1},
        pages = {996-1018},
          doi = {10.1093/mnras/stx2403},
archivePrefix = {arXiv},
       eprint = {1709.06101},
 primaryClass = {astro-ph.GA},
       adsurl = {https://ui.adsabs.harvard.edu/abs/2018MNRAS.473..996M}
}

@ARTICLE{Johnson2016gamma,
       author = {{Johnson}, L. Clifton and {Seth}, Anil C. and {Dalcanton}, Julianne J. and {Beerman}, Lori C. and {Fouesneau}, Morgan and {Lewis}, Alexia R. and {Weisz}, Daniel R. and {Williams}, Benjamin F. and {Bell}, Eric F. and {Dolphin}, Andrew E. and {Larsen}, S{\o}ren S. and {Sandstrom}, Karin and {Skillman}, Evan D.},
        title = "{Panchromatic Hubble Andromeda Treasury. XVI. Star Cluster Formation Efficiency and the Clustered Fraction of Young Stars}",
      journal = {\apj},
         year = 2016,
        month = aug,
       volume = {827},
       number = {1},
          eid = {33},
        pages = {33},
          doi = {10.3847/0004-637X/827/1/33},
archivePrefix = {arXiv},
       eprint = {1606.05349},
 primaryClass = {astro-ph.GA},
       adsurl = {https://ui.adsabs.harvard.edu/abs/2016ApJ...827...33J}
}

@ARTICLE{Fiore2023,
       author = {{Fiore}, Fabrizio and {Ferrara}, Andrea and {Bischetti}, Manuela and {Feruglio}, Chiara and {Travascio}, Andrea},
        title = "{Dusty-wind-clear JWST Super-early Galaxies}",
      journal = {\apjl},
         year = 2023,
        month = feb,
       volume = {943},
       number = {2},
          eid = {L27},
        pages = {L27},
          doi = {10.3847/2041-8213/acb5f2},
archivePrefix = {arXiv},
       eprint = {2211.08937},
 primaryClass = {astro-ph.GA},
       adsurl = {https://ui.adsabs.harvard.edu/abs/2023ApJ...943L..27F}
}

@ARTICLE{Ferrara_2023_monsters,
       author = {{Ferrara}, Andrea and {Pallottini}, Andrea and {Dayal}, Pratika},
        title = "{On the stunning abundance of super-early, luminous galaxies revealed by JWST}",
      journal = {\mnras},
         year = 2023,
        month = jul,
       volume = {522},
       number = {3},
        pages = {3986-3991},
          doi = {10.1093/mnras/stad1095},
archivePrefix = {arXiv},
       eprint = {2208.00720},
 primaryClass = {astro-ph.GA},
       adsurl = {https://ui.adsabs.harvard.edu/abs/2023MNRAS.522.3986F}
}

@INPROCEEDINGS{MORFEO_2024,
       author = {{Ciliegi}, Paolo and {Agapito}, Guido and {Aliverti}, Matteo and {Annibali}, Francesca and {Aridiacono}, Carmelo and {Azzaroli}, Nicol{\`o} and {Balestra}, Andrea and {Baronchelli}, Ivano and {Ballone}, Alessandro and {Baruffolo}, Andrea and {Battaini}, Federico and {Benedetti}, Simone and {Bergomi}, Maria and {Bianco}, Andrea and {Bonaglia}, Marco and {Briguglio}, Runa and {Busoni}, Lorenzo and {Cantiello}, Michele and {Capasso}, Giulio and {Carl{\`a}}, Giulia and {Carolo}, Elena and {Cascone}, Enrico and {Chauvin}, Ga{\"e}l. and {Chebbo}, Manal and {Chinellato}, Simonetta and {Cianniello}, Vincenzo and {Colapietro}, Mirko and {Correia}, Jean-Jacques and {Cosentino}, Giuseppe and {Costa}, Elia and {D'Auria}, Domenico and {De Caprio}, Vincenzo and {Devaney}, Nicholas and {Di Antonio}, Ivan and {Di Cianno}, Amico and {Di Dato}, Andrea and {Di Filippo}, Simone and {Di Francesco}, Benedetta and {Di Giammatteo}, Ugo and {Di Prospero}, Chiara and {Di Rico}, Gianluca and {Di Rocco}, Andrea and {Diretto}, Daphne and {Dolci}, Mauro and {Eredia}, Christian and {Esposito}, Simone and {Fantinel}, Daniela and {Farinato}, Jacopo and {Feautrier}, Philippe and {Foppiani}, Italo and {Genoni}, Matteo and {Giro}, Enrico and {Gluck}, Laurence and {Goncharov}, Alexander and {Grani}, Paolo and {Greggio}, Davide and {Guieu}, Sylvain and {Gullieuszik}, Marco and {Hubert}, Zoltan and {Jocou}, Laurent and {Lampitelli}, Salvatore and {Lapucci}, Tommaso and {Laudisio}, Fulvio and {Leal}, Vincent and {Magnard}, Yves and {Magrin}, Demetrio and {Malone}, Deborah and {Marafatto}, Luca and {Michel}, Christophe and {Mouillet}, David and {Moulin}, Thibaut and {Munari}, Matteo and {Oberti}, Sylvain and {Pancher}, Fabrice and {Pariani}, Giorgio and {Petrella}, Amedeo and {Pinnard}, Laurent and {Plantet}, Cedric and {Portaluri}, Elisa and {Puglisi}, Alfio and {Rabou}, Patrick and {Radhakrishnan}, Kalyan and {Ragazzoni}, Roberto and {Redaelli}, Edoardo Maria Alberto and {Riva}, Marco and {Rochat}, Sylvain and {Rodeghiero}, Gabriele and {Rosignoli}, Luca and {Salasnich}, Bernardo and {Savarese}, Salvatore and {Scalera}, Marcello and {Schipani}, Pietro and {Selvestrel}, Danilo and {Sassolas}, Benoit and {Sordo}, Rosanna and {Teodori}, Ludovico and {Umbriaco}, Gabriele and {Valentini}, Angelo and {Xompero}, Marco},
        title = "{MORFEO at ELT: the adaptive optics module for ELT}",
    booktitle = {Adaptive Optics Systems IX},
         year = 2024,
       editor = {{Jackson}, Kathryn J. and {Schmidt}, Dirk and {Vernet}, Elise},
       series = {Society of Photo-Optical Instrumentation Engineers (SPIE) Conference Series},
       volume = {13097},
        month = aug,
          eid = {1309722},
        pages = {1309722},
          doi = {10.1117/12.3019058},
       adsurl = {https://ui.adsabs.harvard.edu/abs/2024SPIE13097E..22C}
}

@INPROCEEDINGS{MICADO_2024_status,
       author = {{Sturm}, E. and {Davies}, R. and {Alves}, J. and {Cl{\'e}net}, Y. and {Kotilainen}, J. and {Monna}, A. and {Nicklas}, H. and {Pott}, J. -U. and {Tolstoy}, E. and {Vulcani}, B. and {Achren}, J. and {Annadevara}, S. and {Anwand-Heerwart}, H. and {Arcidiacono}, C. and {Barboza}, S. and {Barl}, L. and {Baudoz}, P. and {Bender}, R. and {Bezawada}, N. and {Biondi}, F. and {Bizenberger}, P. and {Blin}, A. and {Bon{\'e}}, A. and {Bonifacio}, P. and {Borgo}, B. and {Born}, J. van den and {Buey}, T. and {Cao}, Y. and {Chapron}, F. and {Chauvin}, G. and {Chemla}, F. and {Cloiseau}, K. and {Cohen}, M. and {Colin}, C. and {Czoske}, O. and {Dette}, J. -O. and {Deysenroth}, M. and {Dijkstra}, E. and {Dreizler}, S. and {Dupuis}, O. and {Egmond}, G. van and {Eisenhauer}, F. and {Elswijk}, E. and {Emslander}, A. and {Fabricius}, M. and {Fasola}, G. and {Ferreira}, F. and {F{\"o}rster Schreiber}, N. and {Fontana}, A. and {Gaudemard}, J. and {Gautherot}, N. and {Gendron}, E. and {Gennet}, C. and {Genzel}, R. and {Ghouchou}, L. and {Gillessen}, S. and {Gratadour}, D. and {Grazian}, A. and {Grupp}, F. and {Guieu}, S. and {Gullieuszik}, M. and {Haan}, M. de and {Hartke}, J. and {Hartl}, M. and {Haussmann}, F. and {Helin}, T. and {Hess}, H. -J. and {Hofferbert}, R. and {Huber}, H. and {Huby}, E. and {Huet}, J. -M. and {Ives}, D. and {Janssen}, A. and {Jaufmann}, P. and {Jilg}, T. and {Jodlbauer}, D. and {Jost}, J. and {Kausch}, W. and {Kellermann}, H. and {Kerber}, F. and {Kravcar}, H. and {Kravchenko}, K. and {Kulcs{\'a}r}, C. and {Kuncarayakti}, H. and {Kunst}, P. and {Kwast}, S. and {Lang}, F. and {Lange}, J. and {Lapeyrere}, V. and {Le Ruyet}, B. and {Leschinski}, K. and {Locatelli}, H. and {Massari}, D. and {Mattila}, S. and {Mei}, S. and {Merlin}, F. and {Meyer}, E. and {Michel}, C. and {Mohr}, L. and {Montarg{\`e}s}, M. and {M{\"u}ller}, F. and {M{\"u}nch}, N. and {Navarro}, R. and {Neumann}, U. and {Neumayer}, N. and {Neumeier}, L. and {Pedichini}, F. and {Pfl{\"u}ger}, A. and {Piazzesi}, R. and {Pinard}, L. and {Porras}, J. and {Portulari}, E. and {Przybilla}, N. and {Rabien}, S. and {Raffard}, J. and {Ragazzoni}, R. and {Ramlau}, R. and {Ramos}, J. and {Ramsay}, S. and {Raynaud}, H. -F. and {Rhode}, P. and {Richter}, A. and {Rix}, H. -W. and {Rodenhuis}, M. and {Rohloff}, R. -R. and {Romp}, R. and {Rousselot}, P. and {Sabha}, N. and {Sassolas}, B. and {Schlichter}, J. and {Schuil}, M. and {Schweitzer}, M. and {Seemann}, U. and {Sevin}, A. and {Simioni}, M. and {Spallek}, L. and {S{\"o}nmez}, A. and {Suuronen}, J. and {Taburet}, S. and {Thomas}, J. and {Tisserand}, E. and {Vaccari}, P. and {Valenti}, E. and {Verdoes Kleijn}, G. and {Verdugo}, M. and {Vidal}, F. and {Wagner}, R. and {Wegner}, M. and {Winden}, D. van and {Witschel}, J. and {Zanella}, A. and {Zeilinger}, W. and {Ziegleder}, J. and {Ziegler}, B.},
        title = "{The MICADO first light imager for the ELT: overview and current status}",
    booktitle = {Ground-based and Airborne Instrumentation for Astronomy X},
         year = 2024,
       editor = {{Bryant}, Julia J. and {Motohara}, Kentaro and {Vernet}, Jo{\"e}l. R.~D.},
       series = {Society of Photo-Optical Instrumentation Engineers (SPIE) Conference Series},
       volume = {13096},
        month = jul,
          eid = {1309611},
        pages = {1309611},
          doi = {10.1117/12.3017752},
       adsurl = {https://ui.adsabs.harvard.edu/abs/2024SPIE13096E..11S}
}

@ARTICLE{Messa_D1T1_2025,
       author = {{Messa}, M. and {Vanzella}, E. and {Loiacono}, F. and {Bergamini}, P. and {Castellano}, M. and {Sun}, B. and {Willott}, C. and {Windhorst}, R.~A. and {Yan}, H. and {Angora}, G. and {Rosati}, P. and {Adamo}, A. and {Annibali}, F. and {Bolamperti}, A. and {Brada{\v{c}}}, M. and {Bradley}, L.~D. and {Calura}, F. and {Claeyssens}, A. and {Comastri}, A. and {Conselice}, C.~J. and {D'Silva}, J.~C.~J. and {Dickinson}, M. and {Frye}, B.~L. and {Grillo}, C. and {Grogin}, N.~A. and {Gruppioni}, C. and {Koekemoer}, A.~M. and {Meneghetti}, M. and {Me{\v{s}}tri{\'c}}, U. and {Pascale}, R. and {Ravindranath}, S. and {Ricotti}, M. and {Summers}, J. and {Zanella}, A.},
        title = "{Anatomy of a z = 6 Lyman-{\ensuremath{\alpha}} emitter down to parsec scales: Extreme UV slopes, metal-poor regions, and possibly leaking star clusters}",
      journal = {\aap},
         year = 2025,
        month = feb,
       volume = {694},
          eid = {A59},
        pages = {A59},
          doi = {10.1051/0004-6361/202451695},
archivePrefix = {arXiv},
       eprint = {2407.20331},
 primaryClass = {astro-ph.GA},
       adsurl = {https://ui.adsabs.harvard.edu/abs/2025A&A...694A..59M}
}

@ARTICLE{Fujimoto_2024_grapes,
       author = {{Fujimoto}, S. and {Ouchi}, M. and {Kohno}, K. and {Valentino}, F. and {Gim{\'e}nez-Arteaga}, C. and {Brammer}, G.~B. and {Furtak}, L.~J. and {Kohandel}, M. and {Oguri}, M. and {Pallottini}, A. and {Richard}, J. and {Zitrin}, A. and {Bauer}, F.~E. and {Boylan-Kolchin}, M. and {Dessauges-Zavadsky}, M. and {Egami}, E. and {Finkelstein}, S.~L. and {Ma}, Z. and {Smail}, I. and {Watson}, D. and {Hutchison}, T.~A. and {Rigby}, J.~R. and {Welch}, B.~D. and {Ao}, Y. and {Bradley}, L.~D. and {Caminha}, G.~B. and {Caputi}, K.~I. and {Espada}, D. and {Endsley}, R. and {Fudamoto}, Y. and {Gonz{\'a}lez-L{\'o}pez}, J. and {Hatsukade}, B. and {Koekemoer}, A.~M. and {Kokorev}, V. and {Laporte}, N. and {Lee}, M. and {Magdis}, G.~E. and {Ono}, Y. and {Rizzo}, F. and {Shibuya}, T. and {Shimasaku}, K. and {Sun}, F. and {Toft}, S. and {Umehata}, H. and {Wang}, T. and {Yajima}, H.},
        title = "{Primordial rotating disk composed of at least 15 dense star-forming clumps at cosmic dawn}",
      journal = {Nature Astronomy},
         year = 2025,
        month = aug,
          doi = {10.1038/s41550-025-02592-w},
archivePrefix = {arXiv},
       eprint = {2402.18543},
 primaryClass = {astro-ph.GA},
       adsurl = {https://ui.adsabs.harvard.edu/abs/2025NatAs.tmp..174F}
}

@ARTICLE{whitmore2010,
       author = {{Whitmore}, Bradley C. and {Chandar}, Rupali and {Schweizer}, Fran{\c{c}}ois and {Rothberg}, Barry and {Leitherer}, Claus and {Rieke}, Marcia and {Rieke}, George and {Blair}, W.~P. and {Mengel}, S. and {Alonso-Herrero}, A.},
        title = "{The Antennae Galaxies (NGC 4038/4039) Revisited: Advanced Camera for Surveys and NICMOS Observations of a Prototypical Merger}",
      journal = {\aj},
         year = 2010,
        month = jul,
       volume = {140},
       number = {1},
        pages = {75-109},
          doi = {10.1088/0004-6256/140/1/75},
archivePrefix = {arXiv},
       eprint = {1005.0629},
 primaryClass = {astro-ph.EP},
       adsurl = {https://ui.adsabs.harvard.edu/abs/2010AJ....140...75W}
}

@ARTICLE{elmegreen_Debra2017,
       author = {{Elmegreen}, Debra Meloy and {Elmegreen}, Bruce G.},
        title = "{Little Blue Dots in the Hubble Space Telescope Frontier Fields: Precursors to Globular Clusters?}",
      journal = {\apjl},
         year = 2017,
        month = dec,
       volume = {851},
       number = {2},
          eid = {L44},
        pages = {L44},
          doi = {10.3847/2041-8213/aaa0ce},
archivePrefix = {arXiv},
       eprint = {1712.02935},
 primaryClass = {astro-ph.GA},
       adsurl = {https://ui.adsabs.harvard.edu/abs/2017ApJ...851L..44E}
}

@ARTICLE{adamo_sparkelr_2023,
       author = {{Adamo}, Angela and {Usher}, Christopher and {Pfeffer}, Joel and {Claeyssens}, Ad{\'e}la{\"\i}de},
        title = "{The ages and metallicities of the globular clusters in the Sparkler}",
      journal = {\mnras},
         year = 2023,
        month = oct,
       volume = {525},
       number = {1},
        pages = {L6-L10},
          doi = {10.1093/mnrasl/slad084},
archivePrefix = {arXiv},
       eprint = {2306.11814},
 primaryClass = {astro-ph.GA},
       adsurl = {https://ui.adsabs.harvard.edu/abs/2023MNRAS.525L...6A}
}

@ARTICLE{ormerod2024_sizes,
       author = {{Ormerod}, K. and {Conselice}, C.~J. and {Adams}, N.~J. and {Harvey}, T. and {Austin}, D. and {Trussler}, J. and {Ferreira}, L. and {Caruana}, J. and {Lucatelli}, G. and {Li}, Q. and {Roper}, W.~J.},
        title = "{EPOCHS VI: the size and shape evolution of galaxies since z   8 with JWST Observations}",
      journal = {\mnras},
         year = 2024,
        month = jan,
       volume = {527},
       number = {3},
        pages = {6110-6125},
          doi = {10.1093/mnras/stad3597},
archivePrefix = {arXiv},
       eprint = {2309.04377},
 primaryClass = {astro-ph.GA},
       adsurl = {https://ui.adsabs.harvard.edu/abs/2024MNRAS.527.6110O}
}

@ARTICLE{krumholz2019,
       author = {{Krumholz}, Mark R. and {McKee}, Christopher F. and {Bland-Hawthorn}, Joss},
        title = "{Star Clusters Across Cosmic Time}",
      journal = {\araa},
         year = 2019,
        month = aug,
       volume = {57},
        pages = {227-303},
          doi = {10.1146/annurev-astro-091918-104430},
archivePrefix = {arXiv},
       eprint = {1812.01615},
 primaryClass = {astro-ph.GA},
       adsurl = {https://ui.adsabs.harvard.edu/abs/2019ARA&A..57..227K}
}

@ARTICLE{Calzetti2000,
       author = {{Calzetti}, Daniela and {Armus}, Lee and {Bohlin}, Ralph C. and {Kinney}, Anne L. and {Koornneef}, Jan and {Storchi-Bergmann}, Thaisa},
        title = "{The Dust Content and Opacity of Actively Star-forming Galaxies}",
      journal = {\apj},
         year = 2000,
        month = apr,
       volume = {533},
       number = {2},
        pages = {682-695},
          doi = {10.1086/308692},
archivePrefix = {arXiv},
       eprint = {astro-ph/9911459},
 primaryClass = {astro-ph},
       adsurl = {https://ui.adsabs.harvard.edu/abs/2000ApJ...533..682C}
}

@ARTICLE{Welch22_clumps,
       author = {{Welch}, Brian and {Coe}, Dan and {Zitrin}, Adi and {Diego}, Jose M. and {Windhorst}, Rogier and {Mandelker}, Nir and {Vanzella}, Eros and {Ravindranath}, Swara and {Zackrisson}, Erik and {Florian}, Michael and {Bradley}, Larry and {Sharon}, Keren and {Brada{\v{c}}}, Maru{\v{s}}a and {Rigby}, Jane and {Frye}, Brenda and {Fujimoto}, Seiji},
        title = "{RELICS: Small-scale Star Formation in Lensed Galaxies at z = 6-10}",
      journal = {\apj},
         year = 2023,
        month = jan,
       volume = {943},
       number = {1},
          eid = {2},
        pages = {2},
          doi = {10.3847/1538-4357/aca8a8},
archivePrefix = {arXiv},
       eprint = {2207.03532},
 primaryClass = {astro-ph.GA},
       adsurl = {https://ui.adsabs.harvard.edu/abs/2023ApJ...943....2W}
}

@ARTICLE{mestric2023,
       author = {{Me{\v{s}}tri{\'c}}, U. and {Vanzella}, E. and {Upadhyaya}, A. and {Martins}, F. and {Marques-Chaves}, R. and {Schaerer}, D. and {Guibert}, J. and {Zanella}, A. and {Grillo}, C. and {Rosati}, P. and {Calura}, F. and {Caminha}, G.~B. and {Bolamperti}, A. and {Meneghetti}, M. and {Bergamini}, P. and {Mercurio}, A. and {Nonino}, M. and {Pascale}, R.},
        title = "{Clues on the presence and segregation of very massive stars in the Sunburst Lyman-continuum cluster at z = 2.37}",
      journal = {\aap},
         year = 2023,
        month = may,
       volume = {673},
          eid = {A50},
        pages = {A50},
          doi = {10.1051/0004-6361/202345895},
archivePrefix = {arXiv},
       eprint = {2301.04672},
 primaryClass = {astro-ph.GA},
       adsurl = {https://ui.adsabs.harvard.edu/abs/2023A&A...673A..50M}
}

@ARTICLE{Mestric22,
       author = {{Me{\v{s}}tri{\'c}}, U. and {Vanzella}, E. and {Zanella}, A. and {Castellano}, M. and {Calura}, F. and {Rosati}, P. and {Bergamini}, P. and {Mercurio}, A. and {Meneghetti}, M. and {Grillo}, C. and {Caminha}, G.~B. and {Nonino}, M. and {Merlin}, E. and {Cupani}, G. and {Sani}, E.},
        title = "{Exploring the physical properties of lensed star-forming clumps at 2 {\ensuremath{\lesssim}} z {\ensuremath{\lesssim}} 6}",
      journal = {\mnras},
         year = 2022,
        month = nov,
       volume = {516},
       number = {3},
        pages = {3532-3555},
          doi = {10.1093/mnras/stac2309},
archivePrefix = {arXiv},
       eprint = {2202.09377},
 primaryClass = {astro-ph.GA},
       adsurl = {https://ui.adsabs.harvard.edu/abs/2022MNRAS.516.3532M}
}

@ARTICLE{he20_ricotti,
       author = {{He}, Chong-Chong and {Ricotti}, Massimo and {Geen}, Sam},
        title = "{Simulating star clusters across cosmic time - II. Escape fraction of ionizing photons from molecular clouds}",
      journal = {\mnras},
         year = 2020,
        month = mar,
       volume = {492},
       number = {4},
        pages = {4858-4873},
          doi = {10.1093/mnras/staa165},
archivePrefix = {arXiv},
       eprint = {2001.06109},
 primaryClass = {astro-ph.GA},
       adsurl = {https://ui.adsabs.harvard.edu/abs/2020MNRAS.492.4858H}
}

@ARTICLE{NUMPY2011,
       author = {{van der Walt}, St{\'e}fan and {Colbert}, S. Chris and
         {Varoquaux}, Ga{\"e}l},
        title = "{The NumPy Array: A Structure for Efficient Numerical Computation}",
      journal = {Computing in Science and Engineering},
         year = 2011,
        month = mar,
       volume = {13},
       number = {2},
        pages = {22-30},
          doi = {10.1109/MCSE.2011.37},
archivePrefix = {arXiv},
       eprint = {1102.1523},
 primaryClass = {cs.MS},
       adsurl = {https://ui.adsabs.harvard.edu/abs/2011CSE....13b..22V}
}

@INPROCEEDINGS{MPDAF2019,
       author = {{Piqueras}, L. and {Conseil}, S. and {Shepherd}, M. and {Bacon}, R. and
         {Leclercq}, F. and {Richard}, J.},
        title = "{MPDAF - A Python Package for the Analysis of VLT/MUSE Data}",
    booktitle = {Astronomical Data Analysis Software and Systems XXVI},
         year = 2019,
       editor = {{Molinaro}, Marco and {Shortridge}, Keith and {Pasian}, Fabio},
       series = {Astronomical Society of the Pacific Conference Series},
       volume = {521},
        month = oct,
        pages = {545},
       adsurl = {https://ui.adsabs.harvard.edu/abs/2019ASPC..521..545P}
}

@ARTICLE{matplotlib2007,
       author = {{Hunter}, John D.},
        title = "{Matplotlib: A 2D Graphics Environment}",
      journal = {Computing in Science and Engineering},
         year = 2007,
        month = may,
       volume = {9},
       number = {3},
        pages = {90-95},
          doi = {10.1109/MCSE.2007.55},
       adsurl = {https://ui.adsabs.harvard.edu/abs/2007CSE.....9...90H}
}

@ARTICLE{Tang2023,
       author = {{Tang}, Mengtao and {Stark}, Daniel P. and {Chen}, Zuyi and {Mason}, Charlotte and {Topping}, Michael and {Endsley}, Ryan and {Senchyna}, Peter and {Plat}, Ad{\`e}le and {Lu}, Ting-Yi and {Whitler}, Lily and {Robertson}, Brant and {Charlot}, St{\'e}phane},
        title = "{JWST/NIRSpec spectroscopy of z = 7-9 star-forming galaxies with CEERS: new insight into bright Ly{\ensuremath{\alpha}} emitters in ionized bubbles}",
      journal = {\mnras},
         year = 2023,
        month = dec,
       volume = {526},
       number = {2},
        pages = {1657-1686},
          doi = {10.1093/mnras/stad2763},
archivePrefix = {arXiv},
       eprint = {2301.07072},
 primaryClass = {astro-ph.GA},
       adsurl = {https://ui.adsabs.harvard.edu/abs/2023MNRAS.526.1657T}
}

@ARTICLE{bouwens_tiny17,
       author = {{Bouwens}, R.~J. and {Illingworth}, G.~D. and {Oesch}, P.~A. and
         {Atek}, H. and {Lam}, D. and {Stefanon}, M.},
        title = "{Extremely Small Sizes for Faint z ̃ 2-8 Galaxies in the Hubble Frontier Fields: A Key Input for Establishing Their Volume Density and UV Emissivity}",
      journal = {\apj},
         year = 2017,
        month = jul,
       volume = {843},
       number = {1},
          eid = {41},
        pages = {41},
          doi = {10.3847/1538-4357/aa74e4},
archivePrefix = {arXiv},
       eprint = {1608.00966},
 primaryClass = {astro-ph.GA},
       adsurl = {https://ui.adsabs.harvard.edu/abs/2017ApJ...843...41B}
}

@ARTICLE{CFE_li18,
       author = {{Li}, Hui and {Gnedin}, Oleg Y. and {Gnedin}, Nickolay Y.},
        title = "{Star Cluster Formation in Cosmological Simulations. II. Effects of Star Formation Efficiency and Stellar Feedback}",
      journal = {\apj},
         year = 2018,
        month = jul,
       volume = {861},
       number = {2},
          eid = {107},
        pages = {107},
          doi = {10.3847/1538-4357/aac9b8},
archivePrefix = {arXiv},
       eprint = {1712.01219},
 primaryClass = {astro-ph.GA},
       adsurl = {https://ui.adsabs.harvard.edu/abs/2018ApJ...861..107L}
}

@ARTICLE{carnall19,
       author = {{Carnall}, Adam C. and {Leja}, Joel and {Johnson}, Benjamin D. and {McLure}, Ross J. and {Dunlop}, James S. and {Conroy}, Charlie},
        title = "{How to Measure Galaxy Star Formation Histories. I. Parametric Models}",
      journal = {\apj},
         year = 2019,
        month = mar,
       volume = {873},
       number = {1},
          eid = {44},
        pages = {44},
          doi = {10.3847/1538-4357/ab04a2},
archivePrefix = {arXiv},
       eprint = {1811.03635},
 primaryClass = {astro-ph.GA},
       adsurl = {https://ui.adsabs.harvard.edu/abs/2019ApJ...873...44C}
}

@ARTICLE{adamo2024,
       author = {{Adamo}, Angela and {Bradley}, Larry D. and {Vanzella}, Eros and {Claeyssens}, Ad{\'e}la{\"\i}de and {Welch}, Brian and {Diego}, Jose M. and {Mahler}, Guillaume and {Oguri}, Masamune and {Sharon}, Keren and {Abdurro'uf} and {Hsiao}, Tiger Yu-Yang and {Xu}, Xinfeng and {Messa}, Matteo and {Lassen}, Augusto E. and {Zackrisson}, Erik and {Brammer}, Gabriel and {Coe}, Dan and {Kokorev}, Vasily and {Ricotti}, Massimo and {Zitrin}, Adi and {Fujimoto}, Seiji and {Inoue}, Akio K. and {Resseguier}, Tom and {Rigby}, Jane R. and {Jim{\'e}nez-Teja}, Yolanda and {Windhorst}, Rogier A. and {Hashimoto}, Takuya and {Tamura}, Yoichi},
        title = "{Bound star clusters observed in a lensed galaxy 460 Myr after the Big Bang}",
      journal = {\nat},
         year = 2024,
        month = aug,
       volume = {632},
       number = {8025},
        pages = {513-516},
          doi = {10.1038/s41586-024-07703-7},
archivePrefix = {arXiv},
       eprint = {2401.03224},
 primaryClass = {astro-ph.GA},
       adsurl = {https://ui.adsabs.harvard.edu/abs/2024Natur.632..513A}
}

@ARTICLE{adamo20extreme,
       author = {{Adamo}, A. and {Hollyhead}, K. and {Messa}, M. and {Ryon}, J.~E. and {Bajaj}, V. and {Runnholm}, A. and {Aalto}, S. and {Calzetti}, D. and {Gallagher}, J.~S. and {Hayes}, M.~J. and {Kruijssen}, J.~M.~D. and {K{\"o}nig}, S. and {Larsen}, S.~S. and {Melinder}, J. and {Sabbi}, E. and {Smith}, L.~J. and {{\"O}stlin}, G.},
        title = "{Star cluster formation in the most extreme environments: insights from the HiPEEC survey}",
      journal = {\mnras},
         year = 2020,
        month = dec,
       volume = {499},
       number = {3},
        pages = {3267-3294},
          doi = {10.1093/mnras/staa2380},
archivePrefix = {arXiv},
       eprint = {2008.12794},
 primaryClass = {astro-ph.GA},
       adsurl = {https://ui.adsabs.harvard.edu/abs/2020MNRAS.499.3267A}
}

\begin{appendix} 
\section{Varying the magnification of the Cosmic Gems arc}
\label{mass_mu}

\begin{figure}
\center
\includegraphics[width=\columnwidth]{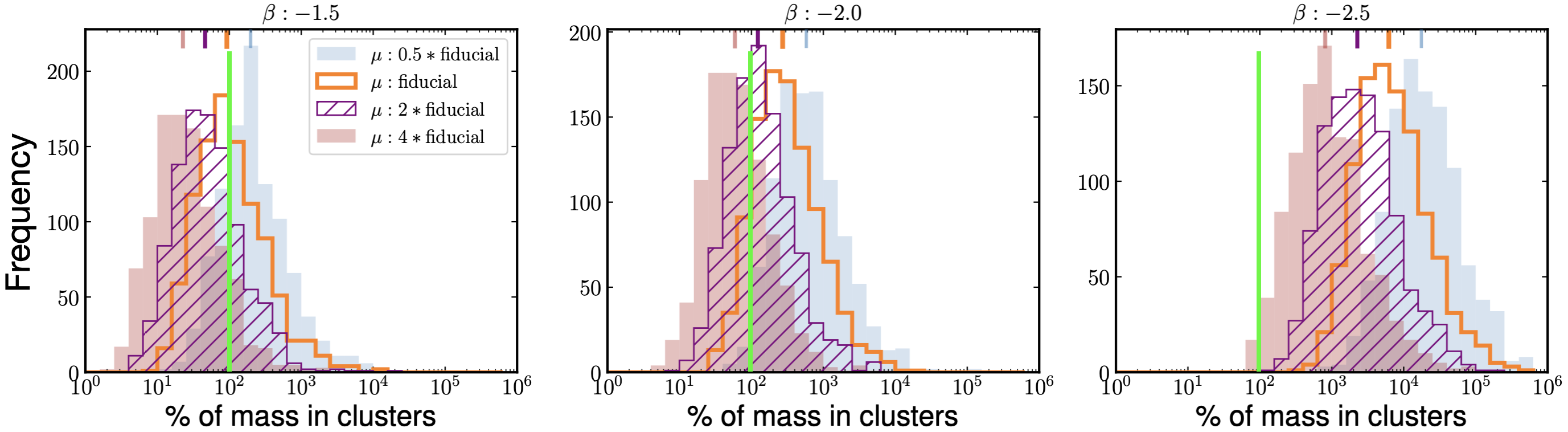}
 \caption{The fraction of the stellar mass of the CG galaxy residing in the bound star cluster population is illustrated. The SCMF is integrated down to the minimum stellar mass of $10^2$~\msun.  Each panel presents four histograms, representing the fraction of stellar mass in star clusters calculated using four different sets of magnifications as indicated in the legend (left panel, and see text for more details). 
 The green vertical line indicates the total stellar mass of the host galaxy located in star clusters.
 }
 \label{multi_hist}
\end{figure}

Three lens models were presented in LB25 all yielding comparable magnifications of the arc. MM25 present new models based on additional multiple systems confirmed in the redshift range $1-6$ with VLT/MUSE spectroscopy (PI. F. Bauer, prog. 0112.A-2069(A)). In this work, we adopt the new fiducial magnification values reported by MM25. The magnifications for the five star clusters including their statistical errors (A, B, C, D, and E) can be written as $V=k \cdot (48^{+4}_{-4}, 92^{+12}_{-11}, 124^{+20}_{-17}, 167^{+32}_{-26}, 323^{+125}_{-82})$, where $k=1.0$ corresponds to the fiducial case.

We then investigate how the results change with variations in $V$. 
Figure~\ref{multi_hist} presents the same quantities as in Figure~\ref{distrib}, adopting three different sets of $V$ corresponding to $k = 0.5, 1.0, 2.0$ and 4.0 (half-fiducial, fiducial, double-fiducial and four times fiducial magnifications, respectively). These values were chosen to explore a wide range of uncertainties\footnote{In more detail, $0.5\leq k \leq2.0$ covers the uncertainty range of the reference lens model used in MM25, while $k=3$ and 4 are included to consider significant deviation from the best fit lens models.} reported in MM25. The magnification gradient along the arc is maintained as suggested by the current fiducial model (MM25), for simplicity. The SCMF is integrated down to a minimum cluster mass of $10^2$ \msun. The magnification of the CI is fixed to $\mu_{host} = 1.84 \pm 0.05$.

Lowering the magnifications below the fiducial values ($k<1$) strengthens the conclusions of this work. Specifically, reduced magnifications (e.g., $k=0.5$) result in a higher mass fraction residing in clusters for any value of $\beta$ adopted in the SCMF, often approaching or even exceeding the stellar mass of the CG galaxy. Conversely, significantly higher magnifications (e.g., $k$=4.0) would be needed to align the stellar mass of the cluster population with values below that of the host galaxy while keeping more relaxed slope of the SCMF and/or minimum stellar mass and/or $\Gamma$. However, it is worth noting that in this case the magnification values are not consistent with any of the lens model predictions described in MM25, and in addition they would imply significantly high stellar mass densities within the star clusters, much higher than those reported by AA24 and MM25. Rather, and more likely, the fiducial magnification values ($k=1$) represent the best scenario, which eventually suggest a possible top-heavy shape of the SCMF, and/or an high M$_{\rm lim}$, and/or a large $\Gamma$ in the CG galaxy.

\section{SED fitting of the counter-image}
\label{SED_FIT}

\begin{figure}
\center
\includegraphics[width=\columnwidth]{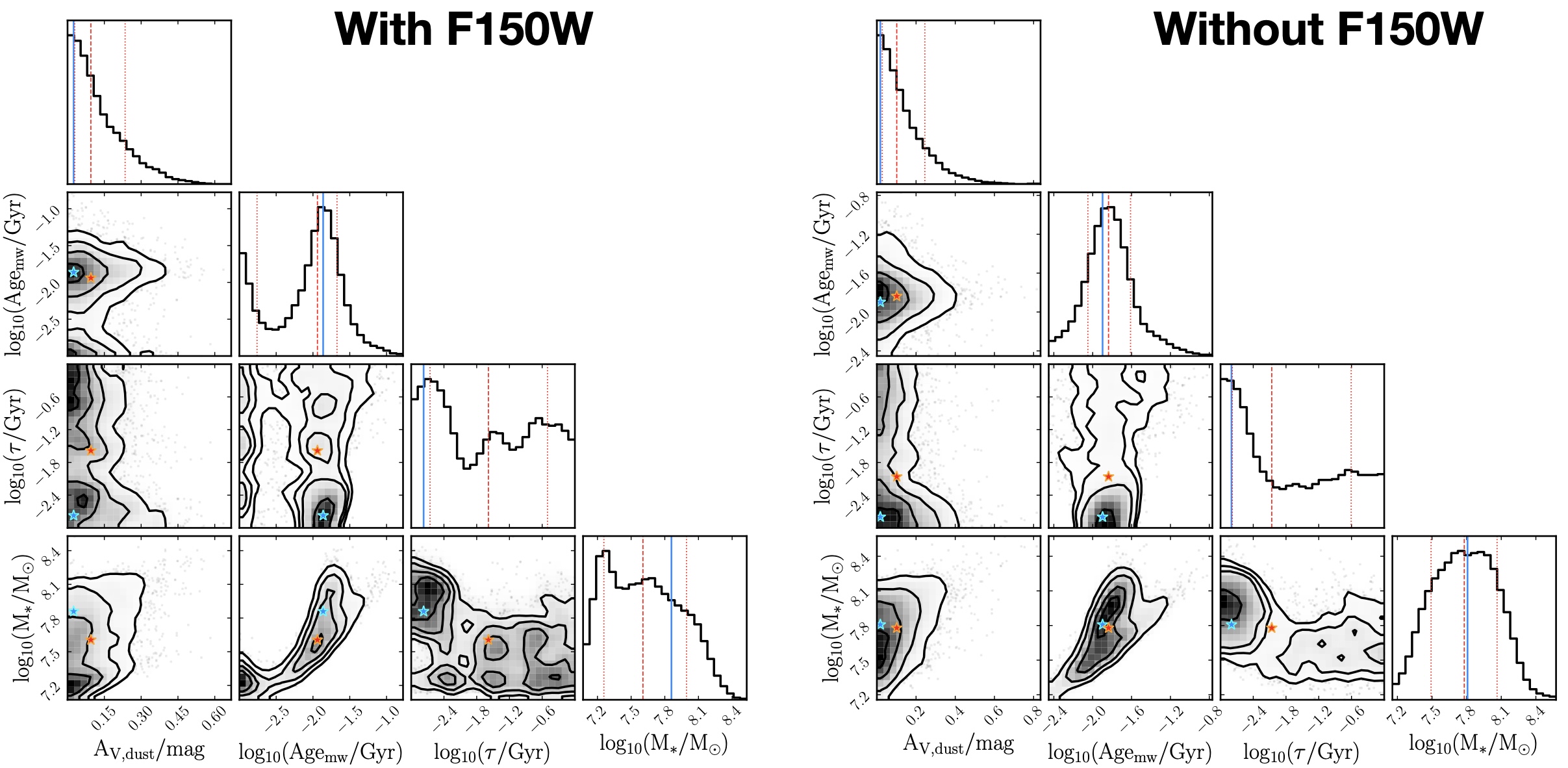}
 \caption{Corner plots for the CI, obtained with and without including the F150W band. Symbols and lines follow the conventions described in Figure~\ref{corner}. The best-fit parameters remain largely consistent overall; however, differences appear in the posterior distributions of $\tau$ and stellar mass distributions (see text for details).
 }
 \label{noF150W}
\end{figure}

Figure~\ref{noF150W} displays the corner plots from the SED fitting of the CI carried out with and without the \JWST/NIRCam F150W band. As shown by MM25 and Christensen et al. in prep., the spectrum exhibits a pronounced \lya\ \emph{damping wing} just redward of the line centre. The resulting flux depression in F150W is clearly visible in Figure~\ref{corner} and matches the deficit measured in the higher–S/N SED of the full arc (LB25). With the spectroscopic redshift now firmly established at $z_{\rm spec}=9.625$, we can understand why the original photometric redshifts were over-estimated, $z_{\rm phot}=10.22\pm0.20$ for the arc and $10.8^{+0.6}_{-1.4}$ for the CI. The fitting algorithm interpreted the F150W attenuation as the onset of the intergalactic \lya\ break sliding into that filter, whereas at $z=9.625$ the break lies blueward of F150W and the flux deficit is instead caused by the intrinsic (or local-CGM) damping wing. Including the F150W point in the fit therefore forces the model SED to bend away from an otherwise consistent solution. Because the exclusion of F150W does not change the best-fit stellar mass or age, we omit this band in the final SED fitting to avoid bias from the damping‑wing absorption.

\end{appendix}

\end{document}